\documentstyle{l-aa}

\def\kort{k_{\perp}}
\def\d{{\rm d}}
\def\mA{{\cal A}^{-1}}
\def\mR{{\cal R}}
\def\mD{{\cal D}}
\def\mU{{\cal U}}

\def\vx{{\bf x}}
\def\vk{{\bf k}}
\def\ii{{\rm i}}
\def\em{\sl}
\def\gam{{\vec{\xi}}}
\def\alf{{\vec{\alpha}}}
\def\mg{\big <}
\def\md{\big >}
\def\Xb{{\overline X}^c}
\def\ovl{\overline}
\def\xib{{\overline\xi}}
\def\wb{{\overline\omega}}

\input psfig.sty

\begin{document}

   \thesaurus{12 (12.04.1; 12.07.1; 12.12.1)} 

 \title{Weak Lensing Statistics as a Probe of $\Omega$ and Power Spectrum}

 \author{F. Bernardeau\inst{1}, L. Van Waerbeke\inst{2}, Y. Mellier\inst{3,4}}

 \offprints{F. Bernardeau; fbernardeau@cea.fr}

 \institute{\inst{1}Service de Physique Th\'eorique, 
C.E. de Saclay, F-91191 Gif-sur-Yvette cedex, France\\
            \inst{2}Observatoire Midi-Pyr\'en\'ees,
14 Av. Edouard Belin, 31400 Toulouse, France.\\
	    \inst{3}Institut d'Astrophysique de Paris, 
98$^{bis}$ Boulevard Arago, 75014 Paris, France.\\
            \inst{4}Observatoire de Paris, (DEMIRM), 61 Av. de l'Observatoire,
75014 Paris, France.}

\maketitle

\markboth{Weak Lensing Statistics}{Bernardeau et al.}

\begin{abstract}

The possibility of detecting weak lensing effects
 from deep wide field imaging surveys
has opened new means of probing the large-scale structure
of the Universe and measuring cosmological parameters. In this paper we
present  a systematic study of the expected dependence
of the low order moments of the filtered 
{\sl gravitational local convergence} on the power spectrum of the density
fluctuations and on the
cosmological parameters $\Omega_0$ and $\Lambda$.
The results show a significant dependence on all these parameters.
Though we note that this degeneracy could be partially 
raised by considering two populations of sources,
at different redshifts,  computing the  third
moment is more promising since it is expected, in the quasi-linear regime 
and for Gaussian initial conditions, to be only $\Omega_0$
dependent (with a slight degeneracy with $\Lambda$) 
when it is correctly expressed in terms of the second moment.

More precisely we show that the variance of the convergence varies 
approximately as 
$P(k) \ \Omega_0^{1.5} \ z_s^{1.5}$, whereas the skewness varies as 
$\Omega_0^{-0.8} \ z_s^{-1.35}$, where $P(k)$ is the projected power
spectrum and $z_s$ the redshift of the sources. Thus, used jointly they can
provide both $P(k)$ and $\Omega_0$. 
However, the dependence on the redshift of
the sources is large and could be a major concern for a practical
implementation.

We have  estimated the errors expected for these parameters in a realistic
scenario and sketched what would be the observational requirements for doing 
such measurements. A more detailed study of an observational
strategy is left for a second paper.
      
\keywords{Cosmology: Dark Matter, Large-Scale Structures, 
Gravitational Lensing}

\end{abstract}

\section{Introduction}

The detection of weak gravitational lensing effects 
(weak shear and magnification bias) at large distance from
cluster
centers is now well established (see Fort \& Mellier 1994 and Narayan \& 
Bartelmann 1996 for reviews). With the present
development of wide field CCD detectors mounted on the best telescopes 
it becomes possible to start the investigation of large scale structures
from lensing effect observations in an angular area of significant
size. The shear and the magnification
are expected to be imposed on the background galaxies by the density
fluctuations along the line-of-sight.
Compared to usual determinations of the density fluctuations
in the local universe with galaxy counts (in fact, the light
distribution), this approach is extremely
attractive since it is directly sensitive to the global
mass fluctuations regardless of any biases associated
with the light distribution.
The detection of weak lensing has  thus naturally
been suggested as a possible means
to measure the power spectrum of the large scale density fluctuations
(Blandford et al. 1991).

Theoretical studies done by Blandford et al. (1991), Miralda-Escud\'e
(1991a), Kaiser (1992) and more recently by Villumsen (1996) proposed various 
approaches and discussed the feasibility of observational strategies 
for some specific cosmological scenarios. In all cases, they concluded 
that weak lensing induced by large scale structures is detectable with
the present techniques.
They claimed that the projected power spectrum should be recovered   
 provided some crucial observational issues, namely the correction of 
 image degradation, are solved. However, even if the observational
aspects are certainly the main difficulties in the future, 
different theoretical problems have not been addressed yet. In 
particular, the dependence
of the physical quantities (such as the convergence, see hereafter)
on the cosmological parameters have 
not been discussed in complete detail. Moreover
 the errors associated with those quantities 
for realistic scenarios of large scale structure formation. 
This paper thus is a theoretical study in which we explore the 
dependence of a priori physical quantities on the cosmological
parameters. We will focus our analysis on the second and third moment of the
local filtered convergence at large angular scale, i.e. in a regime
where Perturbation Theory results are expected to be valid.
We take here advantage of many results that 
have been obtained in the last few years in this regime, showing
that the Perturbation Theory calculations give extremely
accurate results for the behavior of the high order moments
of the cosmic density at large scale (many papers should be quoted,
see Juszkiewicz \& Bouchet 1995 or
Bernardeau 1996b for short review papers on the subject).

Our aim is to present quantitative predictions 
on the behavior of the moments of the one-point
probability distribution function of the local 
convergence as a function of the 
matter power spectrum, the cosmological parameters $\Omega_0$ and $\Lambda$.
As long as one is interested in its second moment in the linear
regime, results can easily be extrapolated to other quantities
such as the magnification or the distortion (see section 3 for more
details on these quantities and their relationships).
For the third moment however, the choice of the local convergence
(or any other scalar quantity) is crucial.
The distortion is intrinsically irrelevant   
since its third moment, for obvious symmetry reasons, should vanish.
The first non-trivial moment would then be its kurtosis, the
connected part of the fourth moment, but this is then more
intricate to calculate, and also would be more difficult
to measure. Other scalar quantities could have been considered, such
as the local magnification, but it is not (at the second
order) directly proportional to the local density, as it is the case for
the convergence. This is why we have preferred this latter
quantity.

The paper is structured as follows. In \S 2 we present the 
basic equations, valid for any cosmological
model, from which we derive the quantities of interest.
In \S 3 we explore the dependence of the variance of the convergence on the
cosmological parameters. In \S 4 we consider third order moments.
In the last section we present a study
of the expected errors due to the use of a finite sample
that are expected to affect the measurements of the second and
third moment. The errors
have been quantified for a realistic scenario (APM like power
spectrum in an Einstein-de Sitter Universe). We eventually propose,
in view of the previous results, an adapted strategy to do such 
observations.
Throughout the paper we adopt the convention $H_0=c=1$, the physical
distances have thus to be multiplied by $3000$ for $H_0=100\  km/s$/Mpc
to be expressed in Mpc.

\section{The physical model for the galaxy and mass distributions}

The gravitational lensing effects
makes intervene both the redshift distributions of the lenses and the sources.
In the last section
we discuss in more detail the knowledge we have or need to have
on the sources, but at this stage we simply describe
the redshift distribution of these objects
by $n(z)\d z$ with the normalization,
\begin{equation}
\int_0^{\infty} \d z\, n(z)=1\ .
\end{equation}
To illustrate the  results we will either assume that 
the sources are all at the same redshift or are given by a broad distribution.
In the latter case we adopt a reasonnable analytic model which reproduces 
the redshift distributions observed in the faint redshift surveys and 
those expected from models of galaxy evolution 
(Charlot \& Fall in preparation),
\begin{equation}
n(z)\d z= {8\over\Gamma(9/8)}
\left({z\over z_s}\right)^8\ \exp\left[-(z/z_s)^8\right]\ {\d z\over z_s}\,,
\label{nz}
\end{equation}
in which it is assumed that the average redshift of the sources
is $z_s$ and  that the width of the distribution
is about $0.4\ z_s$ (see fig. 1, thin lines in the upper panel).

We are interested in the lensing effects induced by the large-scale density
fluctuations that are assumed to have emerged from Gaussian
initial conditions.
In the linear regime we thus assume that the local density can be
written\footnote{Note that in case of a non flat Universe, the product
$\vk\cdot\vx$ should be understood as a covariant product,
$\vk\cdot\vx=\chi k_r+\mD_0\,\kort\cdot\gam$, where $\mD_0$ 
and $\chi$ are respectively the angular distance and the distance
along the line-of-sight (\S 3.3).},
\begin{equation}
\delta(\vx,z)=\int{\d^3 \vk\over (2\ \pi)^{3/2}}
\ D_+(z)\ \delta(\vk)\ e^{\ii \vk\cdot\vx}\label{dk}
\end{equation}
where $D_+(z)$ is the time dependence of the linear growing
mode. Note that the function $D_+(z)$ depends on the cosmological
parameters as well. It is proportional to the expansion factor
only for an Einstein-de Sitter universe.
The Fourier transforms of the initial local density 
contrast, $\delta(\vk)$, are assumed to form a set of Gaussian variables.
In such a case the power spectrum $P(k)$ entirely determines
the statistical properties of the initial density field. It is defined by
\begin{equation}
\mg\delta(\vk)\,\delta(\vk')\md=\delta_{\rm Dirac}(\vk+\vk')
\ P(k).
\end{equation}
From the observed density fluctuation in the APM galaxy
survey it is possible to construct a realistic power spectrum
$P(k)$ for both its magnitude and its shape.
Using the results of Baugh \& Gazta\~naga (1996) 
we can use the shape (we have set  $c=H_0=1$),
\begin{equation}
P(k)=1.2\,10^{-8}{k\over \left[1+(k/k_c)^2\right]^{3/2}},\label{pk}
\end{equation}
with
\begin{equation}
k_c\approx 150,
\end{equation}
that reproduces the observed density fluctuations in the {\sl linear} 
regime\footnote{We are aware that the adopted power spectrum does not have
the expected behaviour, $P(k)\sim k^{-3}$, when $k\to\infty$. But this should
not be a problem since we focus our calculations on the linear or
quasilinear regime. The large $k$ behaviour of $P(k)$
is therefore not relevant for the purpose of this work.}.
 Baugh \& Gazta\~naga  compared  the observed
fluctuations in the APM galaxy
survey with the prediction of this power spectrum 
 and found a good agreement in the linear regime.
Results of numerical simulations also show that the nonlinear evolution
of the density fluctuations induced with such a spectrum are in 
good agreement with the observed full shape of the angular
two-point correlation function.

In this paper, and contrary to previous studies, 
we do not make any assumption on both the density of the
Universe  $\Omega_0$ and on the cosmological constant, $\Lambda$.

\section{The Weak-Lensing Equations}

\subsection{Description of the Deformation in the Image Plane}

The aim of this subsection is to present the mathematical 
objects that describe the local gravitational deformation.
In particular we recall the relationships 
between the local convergence $\kappa$,
the statistical properties of which we will investigate,
and the quantities that are more directly observable.
We consider two neighboring geodesics ${\cal L}$ and
${\cal L}'$ which are enclosed by a light bundle,
and we call $\theta_i$ the two dimensional position angle
between ${\cal L}$ and ${\cal L}'$
at the observer position. Let us define $\alf$ as the angular
diameter distance between ${\cal L}$ and ${\cal L}'$ at a given
redshift $z$. We assume that the small angle approximation is valid, 
and we have,
\begin{equation}
\alpha_i(z)=a\ {\cal D}_{ij}(z)\ \theta_j,\label{map}
\end{equation}
where in the case of an homogeneous Universe, ${\cal D}_{ij}=\mD_0(z)
\delta^K_{ij}$, $a$ being the expansion factor,
$\delta^K_{ij}$ the Kronecker symbol, and $\mD_0(z)$ the physical
distance, (see definition hereafter).
In an inhomogeneous Universe
${\cal D}_{ij}$ describes the deformation of the light bundle produced by
the matter distribution.
The angular distance $\alf$ corresponds to an angular position on the sky
$\gam$ such that,
\begin{equation}
\gam={\alf\over a\,\mD_0}\,.
\end{equation}
The shapes of the light bundles on the image plane are described by
the amplification matrix ${\cal A}$, the Jacobian of
the transformation from a virtual screen located at $z$ (the source
plane), to the image plane as seen by the observer.
Its inverse, $\mA$, is actually more closely related to 
observable quantities, 
\begin{equation}
{\mA}_{ij}(\alf,z)={\partial \xi_i \over \partial \theta_j}\,.
\end{equation}
The amplification matrix can be expressed in terms
of the geometrical deformation of the light bundle, 
the convergence $\kappa$ and the shear components $\gamma_1$ and $\gamma_2$
 (Schneider et al. 1992),
\begin{equation}
{\mA}(\gam)=
\left(
\matrix{1-\kappa-\gamma_1 & -\gamma_2 \cr
-\gamma_2 & 1-\kappa+\gamma_1 \cr}
\right).
\end{equation}
In particular the
local convergence $\kappa(\gam)$ is given by the trace of this
matrix, through the equation,
\begin{equation}
\kappa(\gam)=1-{\rm tr}\left[ {\mA}_{ij}(\gam)\right]/2.\label{kappa}
\end{equation}
The intensity of the shear $\gamma=\gamma_1+i\gamma_2$ is given by the ratio of
the eigenvalues of the matrix $\mA$. 
The relations between the physical quantities $(\kappa,\vec\gamma)$ and the
observable quantities, the magnification (=amplification) $\mu$ and the
distortion $\delta$ of the images, are given by,
\begin{eqnarray}
\mu={1\over |{\mA}|}={1\over (1-\kappa)^2-\gamma^2}\,;\ \ 
\vec\delta={2\vec g\over 1+|g|^2},
\end{eqnarray}
where $\vec g=\vec\gamma/(1-\kappa)$. Note that in the weak lensing
regime we have the simple relations,
\begin{equation}
\mu=1+2\kappa \ ; \ \ \ \vec\delta=2\vec\gamma,
\end{equation}

In practice, the magnification  is measured through the local density of 
objects (Broadhurst et al. 1995, Broadhurst
1995, Fort et al. 1996) or through 
the image size of galaxies in different surface
brightness slices (Bartelmann \& Narayan 1995), whereas  the distortion
is measured directly from the shape of the background galaxies
(Bonnet al. 1994, Fahlman et al. 1994, Smail et al. 1994,
Squires et al. 1996a,b).
One should be careful when order moments of the local 
convergence higher than the variance are considered, since non-linear
couplings exist in general between $\mu$ and $\vec\delta$, and Eqs. (13)
cannot be used.
It is possible to get the convergence from Eqs. (12) by the measurement of
both the magnification and the distortion, or by using only the distortion
and the relation introduced by Kaiser (1995, see also 
Seitz \& Schneider 1996) which is valid even beyond the
linear regime,
\begin{equation}
\vec\nabla\log(1\!-\!\kappa)=
\left(\matrix{1+g_1 & g_2 \cr g_2 & 1-g_1 \cr}\right)^{-1}\cdot
\left(\matrix{ {\displaystyle{\partial g_1\over \partial \theta_1}}
+{\displaystyle{\partial g_2\over \partial \theta_2}}\cr
{\displaystyle{-\partial g_1\over \partial \theta_2}}
+{\displaystyle{\partial g_2\over \partial \theta_1}} }\right).
\end{equation}
In the following we will thus assume that the local 
convergence is accessible to the observations in a given 
sample, and we will focus our investigations on the statistical properties
of this quantity.

\subsection{The Source Equation for the Deformation Matrix}

We perform the calculations assuming that the Born approximation
is valid, i.e. that the deformation of a light bundle can be 
calculated along the unperturbed geodesics.
This assumption will be discussed in section 5.7.
In a Friedmann-Robertson-Walker (FRW) Universe, such a geodesic may be
parameterized by the time variable $\lambda$, defined by,
\begin{equation}
\d \lambda=-{\d a\over H(a)},\ \ \lambda(a=a_0)=0,
\end{equation}
where $H(a)$ is the (time dependent)
Hubble constant (in units of $H_0$).
Using the Born approximation and the geodesic deviation equation,
the evolution equation
for the angular diameter distance $\alf$ in a lumpy Universe, along the
unperturbed ray, as a function of $\lambda$ is (Seitz 1993),
\begin{equation}
\ddot\alf(\lambda)={\cal R}(\lambda)\alf(\lambda),\label{propa}
\end{equation}
where ${\cal R}$ is the $2\times 2$ symmetric tidal matrix, and depends
on the second derivatives of the Newtonian gravitational potential
(Sachs 1961, Seitz 1993),
\begin{eqnarray}
{\cal R}&=&\left(
\matrix{R-{\cal R}e(F) & {\cal I}m(F) \cr
{\cal I}m(F) & R+{\cal R}e(F) \cr}
\right) \ ,\\
R&=&-{1\over 2} R_{\alpha\beta} k^\alpha k^\beta \ ,\\
F&=&-{1\over 2}
C_{\alpha\beta\gamma\delta}\epsilon^{\star\alpha}k^\beta
\epsilon^{\star\gamma} k^\delta\,,
\end{eqnarray}
where $R_{\alpha\beta}$ is the Ricci tensor,
$C_{\alpha\beta\gamma\delta}$ is the Weyl tensor, $k^\alpha$ is the
wave-vector of the light ray
and $\epsilon^\alpha$ is a complex null vector propagated
along the geodesic such that $\epsilon^\alpha k_\alpha=0$.

Let $\varphi$ be the 3-dimensional Newtonian gravitational potential.
The comoving Poisson equation is,
\begin{equation}
\Delta\varphi=4\pi G\ \bar\rho\ \delta,\label{pois}
\end{equation}
where $\delta=(\rho-\bar\rho)/\bar\rho$ is the density contrast, $\Delta$
is the 3-dimensional Laplacian operator, and the
derivatives are done with respect to the proper distance. Note that Eq.
(\ref{pois}) does not depend on the cosmological constant $\Lambda$.
Then, it is straightforward to show that,
\begin{eqnarray}
R&=&
-4\pi G a^{-2} \bar\rho-a^{-2}(\partial_{11}\varphi+\partial_{22}\varphi)\ ,\\
F&=&
-a^{-2}(\partial_{11}\varphi-\partial_{22}\varphi+2i\partial_{12}\varphi).
\end{eqnarray}
From Eqs. (\ref{map}) and (\ref{propa}) the central equation
driving the matrix $\mD$ is then,
\begin{equation}
{\d^2 \left[a(z)\ \mD_{ij}(\gam,z)\right]\over\d \lambda^2}
=a(z)\ \mR_{ik}(\gam,z)\ \mD_{kj}(\gam,z),\label{Eqd}
\end{equation}
with the boundary conditions,
\begin{equation}
\left(\mD_{ij}\right)_{z=0}=0 \ ; \ \ \ \left({\d \mD_{ij}\over \d \lambda}\right)_{z=0}=1\,.
\end{equation}
The first equation expresses the focusing condition at the observer, and
the second the Euclidean properties of the space at small redshift.
$\mR_{ij}$ is a symmetric matrix that can be written in terms of $\varphi$,
and $\bar\rho$ using  the expressions of $R$ and $F$,
\begin{equation}
\mR_{ij}=
-4\pi G\ \bar\rho\ a^{-2}\left(\matrix{1&0\cr0&1}\right)
-2a^{-2}\left(
\matrix{\varphi_{,11} & \varphi_{,12} \cr
\varphi_{,21} & \varphi_{,22} \cr}
\right).
\end{equation}
Since $4\pi G\bar\rho=3/2 \ \Omega_0/a^3$, for a completely
homogeneous universe, the matrix $\mR_{ij}$ is given by, 
\begin{equation}
\mR^{(0)}_{ij}=-{3\over 2} (1+z)^5 \Omega_0 \delta^K_{ij},
\end{equation}
and we are left with the equation for the distance
\begin{equation}
{\d^2 \left[a(z)\ \mD_0(z)\right]\over\d \lambda^2}
=-3/2\ (1+z)^4\ \Omega_0 \mD_0(z)\label{D0}.
\end{equation}
We know that the solution of this differential equation is given by
\begin{equation}
\mD_0(z)={1\over\sqrt{1\!-\!\Omega_0\!-\!\Lambda}}
\sinh\left[\sqrt{1\!-\!\Omega_0\!-\!\Lambda}\int_{0}^z{\d z'\over E(z')}\right],
\end{equation}
with 
\begin{equation}
E[z] = \sqrt{\Lambda+(1+z)^2\ (1\!-\!\Omega_0\!-\!\Lambda)+(1+z)^3\ \Omega_0}.
\end{equation}
This equation has a known analytical solution only when the 
cosmological constant
$\Lambda$ is zero. Otherwise it has to be integrated numerically.

\subsection{The Expression of the Local Convergence}

In this section we restrict ourself to the {\em linear} regime. 
We then assume that the difference between the local value of 
$\mR_{ij}(\gam,z)$ and its value for a homogeneous universe is small, so that,
\begin{eqnarray}
\mR^{(1)}_{ij}(\gam,z)&\approx&
\mR_{ij}(\gam,z)-\mR^{(0)}_{ij}(z)\\
&\equiv& -{3\over 2}\ \Omega_0\ (1+z)^5\ 
\phi^{(1)}_{,ij}(\gam,z),
\end{eqnarray}
where we have introduced an effective potential $\phi$ such that,
\begin{equation}
{1\over 2}\,\Delta\phi^{(1)}(\gam,z)=\delta_{\rm mass}^{(1)}(\gam,z)={\Delta
\varphi^{(1)}\over 4\pi G\bar \rho},
\end{equation}
where $\delta_{\rm mass}^{(1)}(\gam,z)$ is the local mass over-density 
in the linear regime. We assume the small angle deviation approximation
is valid.  In that case, the plane-parallel approximation works, which
means that only waves perpendicular to the line-of-sight contribute to
lensing, and consequently we can neglect the second order derivative
along the unperturbed ray in the Laplacian operator. Moreover,
the difference $\mD_{ij}(\gam,z)-\mD_0(z)\ \delta^K_{ij}$
is a small quantity, which can be expanded with respect
to the initial density. We then define $\mD^{(1)}_{ij}(\gam,z)$
as the part of this difference which is linear in this 
density field (terms of higher order will be considered in 
section 5).
The differential equation for ${\cal D}^{(1)}$ can then be derived
from (\ref{Eqd}) and it reads,
\begin{eqnarray}
&\displaystyle{{\d^2\,\left[a(z)\ \mD^{(1)}_{ij}(\gam,z)\right]\over\d \lambda^2}}
-a(z)\ \mR^{(0)}_{ik}(z)\ \mD^{(1)}_{kj}(\gam,z)=\nonumber\\
&-\displaystyle{{3\over 2}}\ \Omega_0\ (1+z)^4\ \mD_0(z)\ \phi^{(1)}_{,ij}(\gam,z)\label{Eqd1}
\end{eqnarray}
with
\begin{equation}
(\mD^{(1)}_{ij})_{z=0}=0,\ \ \ 
\left({d \mD^{(1)}_{ij}\over d\lambda}\right)_{z=0}=0.
\end{equation}
To solve this differential equation it is more convenient
to write it with the variable $z$. The differential 
equation then reads,
\begin{eqnarray}
&\displaystyle{{\d^2\,\mD^{(1)}_{ij}(\gam,z)\over\d z^2}
+{1\over E(z)}{\d E(z)\over\d z}\,
{\d\mD^{(1)}_{ij}(\gam,z)\over\d z}}-\nonumber\\
&\displaystyle{{1\over 1+z}\,{1\over E(z)}{\d E(z)\over \d z}\,\mD^{(1)}_{ij}(\gam,z)
+{3\over2}{\Omega_0\,(1+z)\over E^2(z)}\,\mD^{(1)}_{ij}(\gam,z)} \\
&=\displaystyle{-{3\over2}{\Omega_0\,(1+z)\over E^2(z)}\,\mD_0(\gam,z)\,
\phi^{(1)}_{,ij}(\gam,z)}.
\end{eqnarray}
The  homogeneous differential
equation associated with this equation has two known solutions,
the distance functions $\mD_0(z)$ and $\mU_0(z)$
 are given respectively by Eq. (28) and,
\begin{eqnarray}
\mU_0(z)={1\over\sqrt{1\!-\!\Omega_0\!-\!\Lambda}}
\cosh\left[\sqrt{1\!-\!\Omega_0\!-\!\Lambda}
\int_{0}^z{\d z'\over E(z')}\right], 
\end{eqnarray}
from which the solution of the differential equation (\ref{Eqd1}) can be
written.
Indeed the solution of the inhomogeneous differential equation
reads
\begin{eqnarray}
&\displaystyle{\mD^{(1)}_{ij}(\gam,z)=-{3\over 2}\,\Omega_0
\int_0^z\,\d z'\,{(1+z')^4\ \mD_0(z')\,\phi^{(1)}_{,ij}(z')\over E^2(z')}}
\times\nonumber\\
&\displaystyle{{\mU_0(z)\,\mD_0(z')-\mU_0(z')\,\mD_0(z)\over
\mU'_0(z')\,\mD_0(z')-\mU_0(z')\,\mD'_0(z')}},
\end{eqnarray}
which, after elementary mathematical transformations simplifies in,
\begin{eqnarray}
&\displaystyle{\mD^{(1)}_{ij}(\gam,z)=-{3\over 2}\ \Omega_0\ 
\int_0^z {\d z'\over E(z')}{1\over\sqrt{1\!-\!\Omega_0\!-\!\Lambda}}}\times\\
&\displaystyle{\sinh\left[\sqrt{1\!-\!\Omega_0\!-\!\Lambda}
\int_{z'}^z{\d\ z''\over E(z'')}\right]
(1+z')\ \mD_0(z')\ \phi^{(1)}_{,ij}}.\nonumber
\end{eqnarray}
It follows from Eq. (\ref{kappa}) that the local convergence can be written
for a source at redshift $z$ (see also Seljak 1996),
\begin{eqnarray}
 &\displaystyle{\kappa^{(1)}(\gam)=-{3\over 2}\ \Omega_0\ 
\int_0^{\chi(z)} {\d\chi(z')}}\times\label{kappa1}\\
&\displaystyle{{\mD_0(z',z)\ \mD_0(z')\ \over \mD_0(z)}
\ (1+z')\ \delta_{\rm mass}^{(1)}(\gam,z')},\nonumber
\end{eqnarray}
where $\mD_0(z',z)$ is the distance between the 
redshifts $z'$ and $z$. We have also introduced the new variable $\chi(z)$,
very useful for such calculations, which is the physical
distance along the line-of-sight, 
\begin{equation}
d \chi(z)={\d z\over E(z)}.
\end{equation}
It coincides with $\mD_0(z)$ only for a flat geometry
$\Omega_0+\Lambda=1$.
Eq. (\ref{kappa1}) expresses the fact that the local
convergence $\Delta\mu=2\kappa^{(1)}$ is given by the superposition
of the convergence induced by each lens between the observer and the source.
For the distribution of sources $n(z)$ we thus have
\begin{eqnarray}
\displaystyle{\kappa^{(1)}(\gam)}=&\displaystyle{-
\int_0^{\infty}{\d z'\over E(z')}\,w(z')}
\displaystyle{\delta_{\rm mass}^{(1)}(\gam,z')},\label{kap11}
\end{eqnarray}
with
\begin{eqnarray}
\displaystyle{w(z')}=&\displaystyle{{3\over 2}\,\Omega_0
\ (1+z')\ \mD_0(z')\ \int_{z'}^{\infty}{\d z\ n(z)\over D_0(z)}\ 
}\times\label{Eff}\\
&\displaystyle{{1\over\sqrt{1\!-\!\Omega_0\!-\!\Lambda}}\,
\sinh\left[\sqrt{1\!-\!\Omega_0\!-\!\Lambda}\int_{z'}^z{\d z''\over E(z'')}\right]}.\nonumber
\end{eqnarray}
The previous two equations are the basic ones
that relate the local convergence to 
the linear cosmic density along the line-of-sight
for any cosmological model.

\subsection{The Efficiency Function}

\begin{figure}
\vspace{11 cm}
\special{hscale=80 vscale=80 voffset=110 hoffset=10 psfile=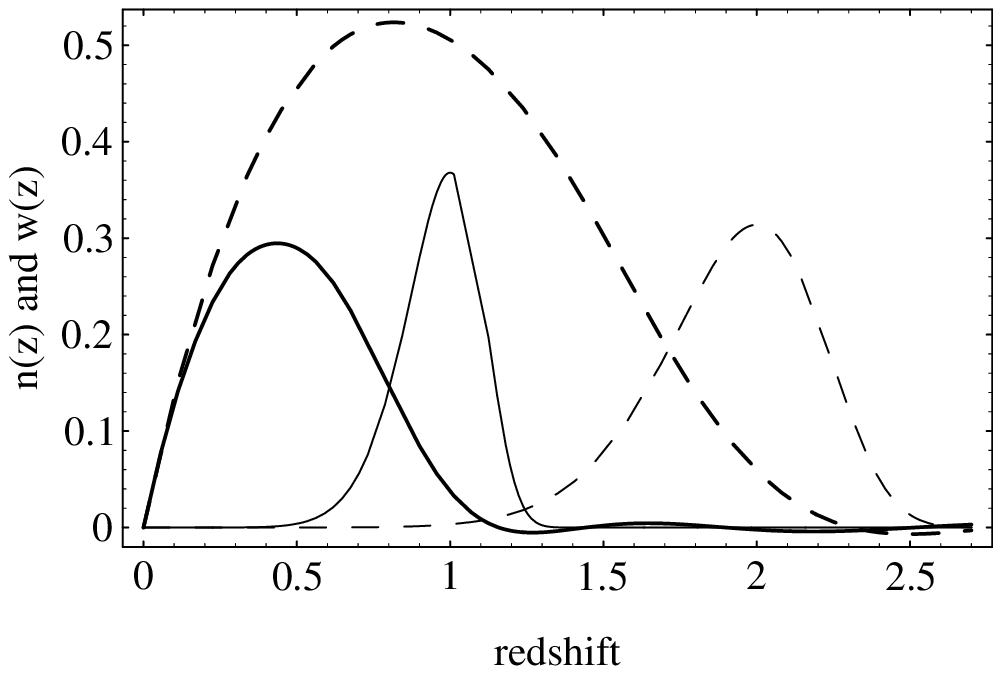}
\special{hscale=80 vscale=80 voffset=-50 hoffset=10 psfile=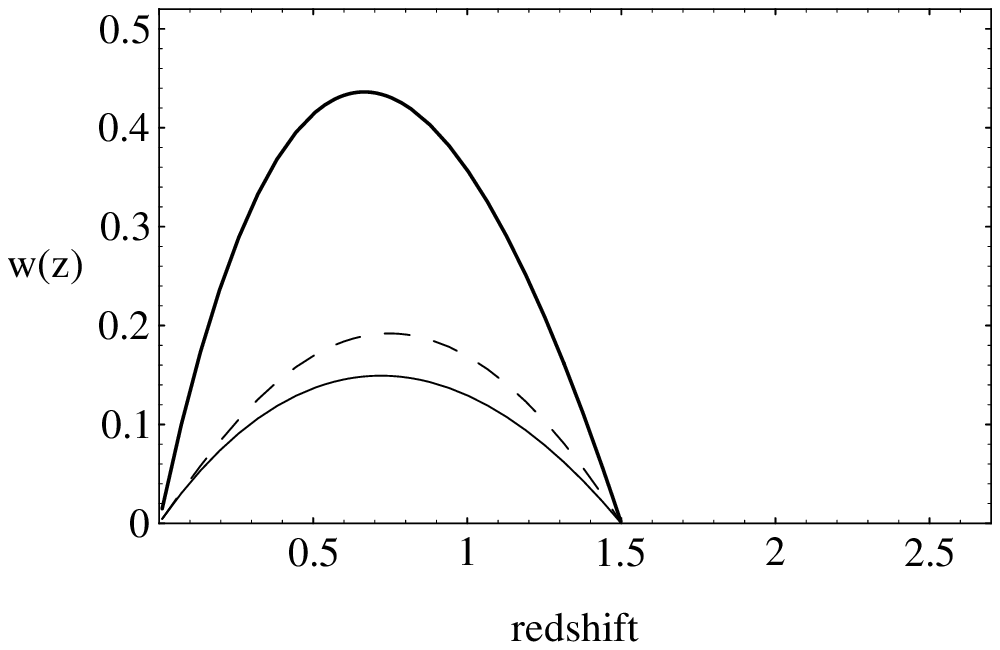}
\caption{Shapes of the efficiency functions in different cases.
In the upper panel the selection functions $n(z)$ are given
by the thin lines (corresponding to Eq. [\ref{nz}] for $z_s=1$
and $z_s$=2) and we assume an Einstein-de Sitter Universe. In the
lower panel the sources are assumed to be all located at $z_s=1.5$
and the cosmological parameters are varied, thick solid line for
$\Omega=1$, $\Lambda=0$, thin solid line for $\Omega=0.3$ $\Lambda=0$ 
and thin dashed line for $\Omega=0.3$ and $\Lambda=0.7$. }
\end{figure}

We call $w(z)$ in Eq. (\ref{Eff}) the efficiency function. It describes
the efficiency with which a lens at a given redshift $z$ located along
the line-of-sight will affect the local convergence. This
efficiency function obviously depends on the redshift of the sources.
It is maximum at half the distance between the source and the observer,
and vanishes at both ends. In fig. 1, we present the shape of the
efficiency function in different case. In 1a, this is for an
Einstein-de Sitter Universe for two different populations of sources.
The thin lines show the shape of the distribution functions of the 
redshifts of the sources that are assumed to be either centered 
on $z=1$ (solid line) or $z=2$ (dashed line). The typical redshift of the 
lens is about 0.4 to 0.5, with a broader distribution when the
redshift of the sources is larger.

In the lower panel we show the dependence of the efficiency function 
on the cosmological parameters. Here the sources are simply
assumed to all lie at $z=1.5$. We can see that the shape
of the efficiency function is not really affected. The amplitude
is however changed, and it is roughly proportional to $\Omega_0$.
The dependence on $\Lambda$, although not totally negligible, is
considerably weaker than that on $\Omega_0$.

\section{The Variance of the Smoothed Convergence}

We are interested in the statistical properties of the local
convergence at a given scale. We are more particularly interested
in the variance at a scale corresponding to the linear
or the quasi-linear
regime where the interpretation in terms of the cosmological
models is easier. For the usual cases, sources at a redshift of about unity,
it corresponds to angular scales of about $30 \arcmin$, or more.

\subsection{The Convolution with the Window Function}

Throughout this paper we assume that the geometrical filtering 
is done with a top-hat window function of angular scale $\theta_0$.
In the expression (\ref{kap11}) 
the smoothing applies to the $\exp({\ii\vk\cdot\vx})$
factor and it gives in the small angle approximation (see Bernardeau 1995),
\begin{eqnarray}
&\displaystyle{{1\over \pi\,\theta_0^2}
\int_0^{\theta_0}\d\theta\int_0^{2\pi}
\d\phi\,\exp(\ii\vk\cdot\vx)}=\label{sm}\\
&\displaystyle{W_{2D}[\kort\mD_0(z)\theta_0]\ 
\exp^{\ii\,k_r\chi(z)}}.\nonumber
\end{eqnarray}
where $k_r$ is the component of $\vk$
along the line-of-sight, $\kort$ its component in the transverse directions
and $W_{2D}$ is the Fourier transform of the 2D top-hat window function,
\begin{equation}
W_{2D}(k)=2{J_1(k)\over k},\label{w2d}
\end{equation}
($J_1$ is the Bessel function of the first kind).
Note that the expression (\ref{sm}) involves both the distance 
$\mD_0(z)$ and the variable $\chi(z)$. These two quantities are identical 
only for cosmological models with zero curvature. 

\subsection{The Variance in the Linear Regime}

Using the definition (\ref{dk}) of the power spectrum  and the relations 
(\ref{sm}, \ref{kap11}) we get
\begin{eqnarray}
&\displaystyle{\mg\kappa^2(\theta_0)\md={1\over 2\pi}
\int_0^{\infty}{\d z_1\over E(z_1)}\int_0^{\infty}{\d z_2\over E(z_2)}}
\times\\
&\displaystyle{w(z_1)\,w(z_2)\,{D_+(z_1)\,D_+(z_2)}
\int{\d^3 \vk\ \over (2\pi)^3}\ 
P(k)}\times\nonumber\\
&\ \displaystyle{W_{2D}^2[\kort\,D_0(z)\,\theta_0]\ 
\exp\left[\ii\,k_r\,(\chi(z_1)\!-\!\chi(z_2))\right]}.\nonumber
\end{eqnarray}
where $D_+(z)$ is the growing mode of the density contrast
(cf. Eq. [3]).

We use again the assumption that the deflecting angle 
and the smoothing angle are small.  It implies that  
\begin{equation}
P(k)\approx P(\kort).
\end{equation}
Then the integral over $k_r$ yields a Dirac delta function in
$\chi(z_1)-\chi(z_2)$ leading to the expression,
\begin{eqnarray}
&\displaystyle{\mg\kappa^2(\theta_0)\md={1\over 2\,\pi}
\int_0^{\infty}{\d z\over E(z)}\ w(z)^2
{D_+^2(z)}}\times\nonumber\\
&\displaystyle{\int_0^{\infty}
\kort\ \d \kort\ P(\kort)\ W_{2D}^2[\kort\,D_0(z)\,\theta_0]}.\label{w2}
\end{eqnarray}
In the following section we explore the properties of this
function with different hypotheses for the power spectrum, the redshift
distribution of the sources, and the cosmological parameters.

\subsection{Realistic Results for an Einstein-de Sitter Universe}

For an Einstein-de Sitter Universe, the distance takes a simple 
expression as a function of the redshift,
\begin{equation}
\mD_0(z)=2-2/\sqrt{1+z}
\end{equation}
and we have,
\begin{eqnarray}
w(z')=&{3\over 2}\,\Omega_0\,(1+z')\,
\mD_0(z')\,\times\nonumber\\
&\int_{z'}^{\infty}\d z\ n(z)\ \left[1-\mD_0(z')/\mD_0(z)\right].
\end{eqnarray}


\begin{figure}
\vspace{5 cm}
\special{hscale=80 vscale=80 voffset=-40 hoffset=10 psfile=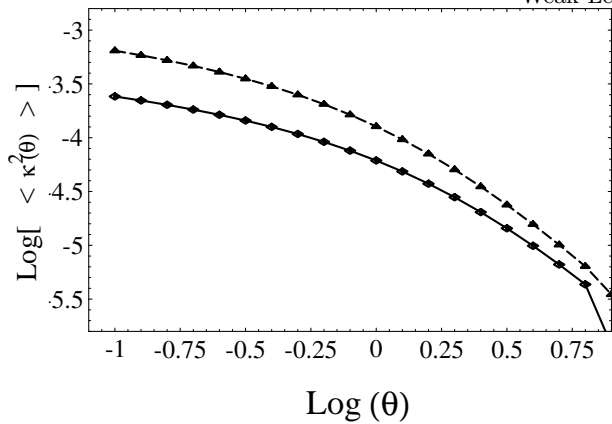}
\caption{The expected variance of the convergence as a function
of the smoothing angle (in degrees) for $z_s=1$ (solid line) and $z_s=2$ 
(dashed line) in (\ref{nz}) and assuming an Einstein-de Sitter Universe.}
\end{figure}

We also know that the linear growth rate of the fluctuation
is proportional to the expansion factor so that 
\begin{equation}
D_+(z)=1/(1+z).
\end{equation}
Using the power spectrum given in (\ref{pk}) and the redshift
distribution (\ref{nz}) for $z_s=1$ and $z_s=2$ we have calculated the
angular dependence of the variance of the local smoothed
convergence. The results are presented in Fig. 2. We can see that 
the typical amplitude of the convergence is of the order of
a few percent. It starts to bend at an angular scale of about 5 to 10
degrees corresponding to the physical scale where the power spectrum
bends down. This scale will be of crucial importance for the evaluation
of the cosmic errors on the measured moments. 

We can also notice that the results have a significant dependence on
the adopted redshift distribution of the sources. In the next 
subsection we discuss in more detail this point.

\subsection{Dependence with the redshift of the Sources}

We explore here the dependence of the second moment
with the redshift of the sources
assuming that they are all at a given redshift $z_s$.
We also assume that the power spectrum is given by a power law
spectrum.
For an Einstein-de Sitter Universe the result reads,
\begin{eqnarray}
&\displaystyle{\mg\kappa^2(\theta_0)\md\propto \int_0^{\mD_0(z_s)}\d 
\mD\,\mD^2\,(1-\mD/\mD_0(z_s))^2\,\mD^{-2-n}}=\nonumber\\
&\displaystyle{{2\over (1-n)(2-n)(3-n)}
\mD_0^{1-n}(z_s)}
\end{eqnarray}
where the angular distance $\mD_0$ is given in the equation (\ref{D0}).
For sources of redshift of the order of 1 or 2, and for a power law 
index of $n\approx-1.5$, we have approximately,
\begin{equation}
\mg\kappa^2(\theta_0)\md\propto z_s^{1.5}.
\end{equation}


This result confirms the trend observed in the
previous subsection. It already indicates that the redshift dependence
of the second moment is rather large. Therefore any interpretation
of a measurement of such a moment in terms of magnitude of the power 
spectrum would require a precise knowledge of the sources
that have been used. Moreover the magnitude of the second moment depends
as well on the values of the cosmological parameters $\Omega_0$
and $\Lambda$. The dependence of the results on those parameters
is explored in the next subsection.

\subsection{Dependence on the Cosmological Parameters}

\begin{figure}
\vspace{7 cm}
\special{hscale=80 vscale=80 voffset=-20 hoffset=10 psfile=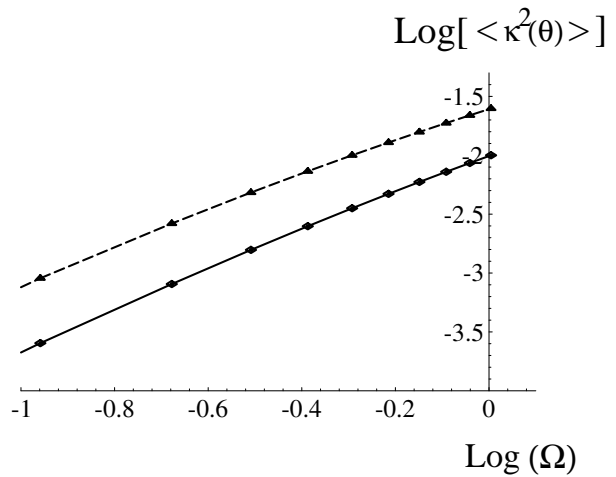}
\caption{The expected dependence of the magnitude of the variance on
$\Omega_0$ for $z_s=1$ (solid line) and $z_s=2$ (dashed line). 
The overall normalization is arbitrary.}
\end{figure}

\begin{figure*}
\vspace{7 cm}
\special{hscale=70 vscale=70 voffset=-10 hoffset=20 psfile=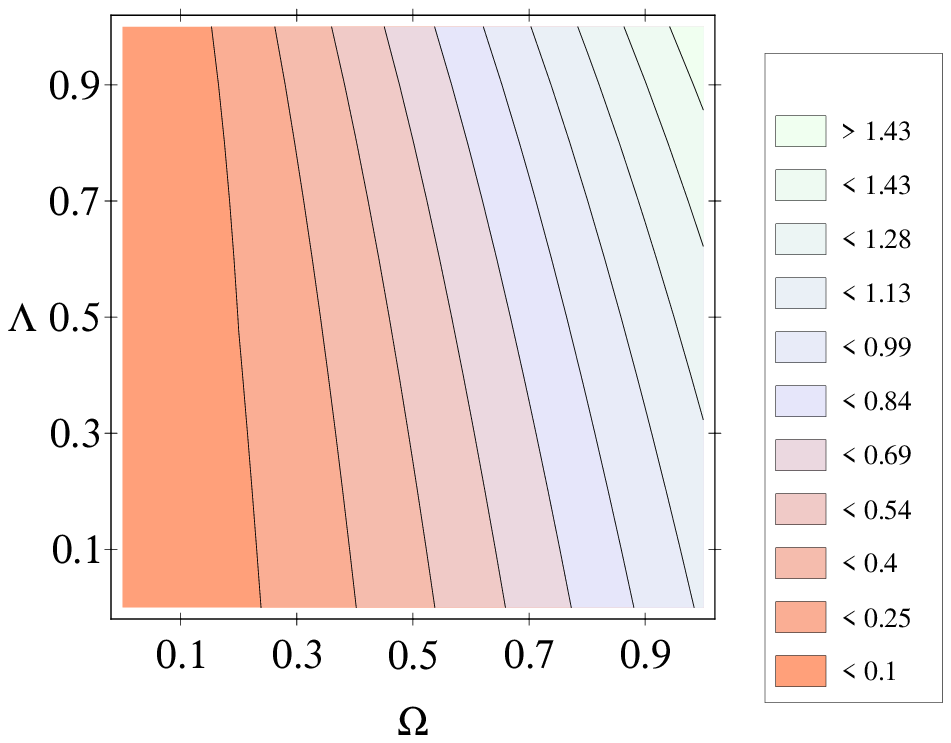}
\special{hscale=70 vscale=70 voffset=-10 hoffset=250 psfile=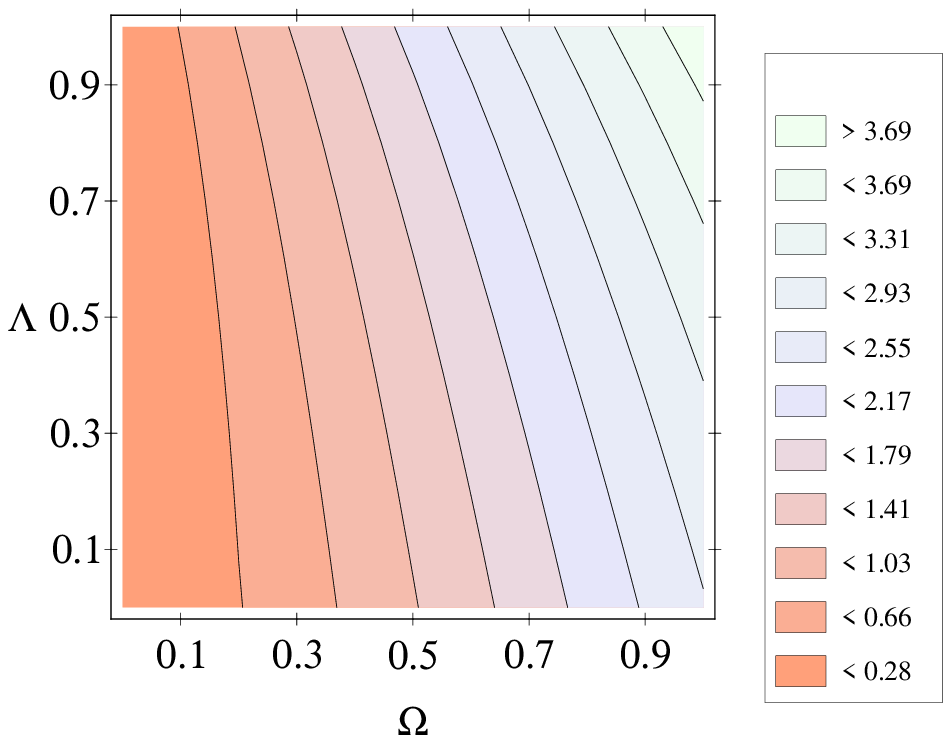}
\caption{Contour plots of 
the expected dependence of the magnitude of the variance on
$\Omega_0$ (horizontal axis) and $\Lambda$ (vertical axis) for
$z_s=1$ (left panel) and $z_s=2$ (right panel). The lines are iso-values
of the variance, normalized to the Einstein-de Sitter case and
regularly spaced in linear scale. The variance is
stronger at the top right corner of the panels.}
\end{figure*}

In this subsection we assume that the power spectrum is a power law and that
the sources are all at redshift $z_s=1$ or $z_s=2$.
In Fig. 3 we then plot the dependence of the second
moment on $\Omega_0$ for $\Lambda=0$.
We have approximately 

\begin{equation}
\mg\kappa^2(\theta_0)\md\propto \Omega_0^{1.6}
\end{equation}
for $z_s=1$ and
\begin{equation}
\mg\kappa^2(\theta_0)\md\propto \Omega_0^{1.4}
\end{equation}
for $z_s=2$. The resulting $\Omega_0$ dependence
is thus weaker than what one would naively expect from Eq. (\ref{w2})
(e.g. $\mg\kappa^2(\theta_0)\md\propto \Omega_0^{2}$).
The fact that the $\Omega_0$ dependence is actually weaker is mainly due
to the growth factor of the linear mode. Indeed at high redshift,
for a given normalization at $z=0$, the lower $\Omega_0$ the 
larger the density fluctuations are.

In Fig. 4 we present a contour plot of the $\Omega_0$ and $\Lambda$
dependence of the second moment.
We can see that the expected magnitude of the second moment depends
essentially on $\Omega_0$. It is however degenerate with $\Lambda$
when $\Omega_0$ is large. As expected, the $\Lambda$ dependence is 
more important when the redshift of the sources is larger.

\subsection{The Variance with two Populations of Sources}

Interestingly when one considers jointly two different populations 
of sources, here with $z_s=1$ and $z_s=2$, the ratio of the two
variances depends on $\Omega_0$ and $\Lambda$ (see contour
plot of Fig. 5) but independent of the normalization of the power
spectrum. (The dependence on the power law index is weak).

This is thus, a priori, a way to disentangle the different parameters.

\section{Higher-Order Moments}

In this section we explore the possibility of considering higher order moments
of the distribution of the local convergence. 
The idea is that the higher order moments are sensitive to 
nonlinear aspects of the dynamics and thus could provide independent 
constraints on the cosmological parameters.

The nonlinear terms in the expression of the fields can be calculated using
Perturbation Theory.
Similar calculations have been
done for the density field and for the local velocity
divergence. Comparisons with numerical results have shown that 
Perturbation Theory calculations give excellent quantitative 
predictions of the behavior of the high order moments
(see Bernardeau 1996b and references therein).

\subsection{Second Order Perturbation Theory for the Local Convergence}

The basic assumption of perturbation theory is  that the local convergence
can be expanded in terms of the initial density field,
\begin{equation}
\kappa(\gam,z)=\kappa^{(1)}(\gam,z)+\kappa^{(2)}(\gam,z)+\dots\label{ExpKa}
\end{equation}
where $\kappa^{(1)}(\gam,z)$ is linear in the initial density
field, so in the variables $\delta(\vk)$
(this is the term calculated previously), and $\kappa^{(2)}(\gam,z)$
is quadratic in those variables, etc...
For calculating the expression of $\kappa^{(2)}(\gam,z)$
we need to use the differential equation (\ref{Eqd}) up to its second
order introducing the expression of 
\begin{equation}
\mR^{(2)}_{ij}(\gam,z)=-{3\over2} \Omega_0\ (1+z)^5\ \phi_{,ij}^{(2)}(\gam,z)
\end{equation}
with 
\begin{equation}
{1\over 2}\Delta \phi^{(2)}(\gam,z)=\delta_{\rm mass}^{(2)}(\gam,z).
\end{equation}
To find the expression of the local convergence at the second order
it is thus necessary to have the expression of the local density
at the same order.

The calculation of the second order term for the local density
has been investigated in many papers. It has now been calculated 
for any cosmological model, and a very useful description is given by,
\begin{eqnarray}
&\displaystyle{\delta_{\rm mass}^{(2)}(\vx)=D_+^2(z)
\int\d^3\vk_1\d^3\vk_2\delta(\vk_1)\delta(\vk_2)}
\times\nonumber\\
&
\displaystyle{\left({5\over7}+{\vk_1\cdot \vk_2\over k_1^2}+
{2\over7}{(\vk_1\cdot \vk_2)^2\over k_1^2 k_2^2}\right)
\exp\left[\ii\vx\cdot(\vk_1+\vk_2)\right]},
\end{eqnarray}
where the $\delta(\vk)$ factors are the Fourier transform of the initial
density field (cf. [3]). 

The coefficients appearing in this expression given for
an Einstein-de Sitter universe, are in fact slightly $\Omega_0$
and $\Lambda$ dependent (Bouchet et al. 1992, Bernardeau 1994). 
These dependences are actually so weak (less than 1\%) that
they can be safely neglected in the subsequent calculations.

The differential equation for $\mD^{(2)}_{ij}(\gam,z)$ can be obtained from
the equation (\ref{Eqd}), when it is developed up to its second order. 
We then get,
\begin{eqnarray}
&\displaystyle{{\d^2\,\left[a(z)\ \mD^{(2)}_{ij}(\gam,z)\right]\over\d \lambda^2}
-a(z)\ \mR^{(0)}_{ik}(z)\ \mD^{(2)}_{kj}(\gam,z)}=\nonumber\\
&\displaystyle{-{3\over 2}\ \Omega_0\ (1+z)^4\ \phi^{(2)}_{ij}(\gam,z)\ \mD_0(z)}+\nonumber\\
&+a(z)\ \mR^{(1)}_{ik}(\gam,z)\ \mD^{(1)}_{kj}(\gam,z)
\end{eqnarray}
with
\begin{equation}
(\mD^{(2)}_{ij})_{z=0}=0\,;\ \ 
\left({d \mD^{(2)}_{ij}\over d\lambda}\right)_{z=0}=0.
\end{equation}
The expression of the local convergence can be deduced from the previous
two equations,
\begin{eqnarray}
&\displaystyle{\kappa^{(2)}(\gam)=-
\int_0^{\infty} {\d z'\over E(z')}\ w(z')\times}\label{Kap2}\\
&\displaystyle{\left[\delta_{\rm mass}^{(2)}(\gam,z')-{1\over 2}{\rm tr}
\left[\phi_{ik}^{(1)}(\gam,z')
\mD^{(1)}_{kj}(\gam,z')\right] /\mD_0(z')\right]}.\nonumber
\end{eqnarray} 
As it can be observed two terms are contributing to this expression.
One term is given by the second order density 
field, the other one is a combination of the linear order of the 
local density and the linear order in the local amplification. 
The latter corresponds to the nonlinear coupling introduced by
two subsequent deflecting lenses. In the following we will
see that the contribution to the third moment is dominated by
the first contribution.

\begin{figure}
\vspace{7 cm}
\special{hscale=80 vscale=80 voffset=-10 hoffset=10 psfile=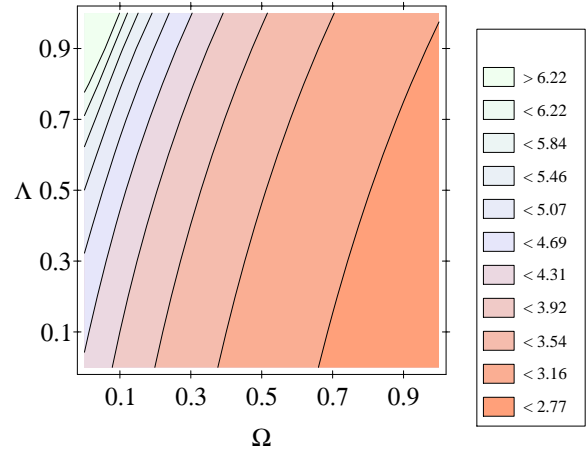}
\caption{Contour plot of 
the expected dependence of the magnitude of the ratio
of the two variances on
$\Omega_0$ (horizontal axis) and $\Lambda$ (vertical axis).}
\end{figure}

\subsection{The Expression of the Local Skewness}

The principle of such a calculation has been developed in previous
papers for the density or the velocity field.
It relies on the assumption that the initial density field
is Gaussian. As a consequence we have,
\begin{equation}
\mg\delta(\vk_1)\dots\delta(\vk_{2p+1})\md=0
\end{equation}
and
\begin{equation}
\mg\delta(\vk_1)\dots\delta(\vk_{2p})\md=\sum_{\rm permutations}
\prod_{j=1}^p\ \mg\delta(\vk_{2j-1}) \delta(\vk_{2j})\md
\end{equation}
where the sum over the permutations is made in such a way that all
possible pair associations are taken into account.

Then, applying those properties to the initial density field
one can calculate the leading term for the expression of the third
moment of the local convergence. Using the expansion (\ref{ExpKa}) we have
\begin{eqnarray}
\mg\kappa^3(\theta_0)\md&=&
\mg\left(\kappa^{(1)}+\kappa^{(2)}+\dots\right)^3\md\nonumber\\
&=&\mg\left(\kappa^{(1)}\right)^3\md+3\ 
\mg\left(\kappa^{(1)}\right)^2\kappa^{(2)}\md+\dots\label{skew}
\end{eqnarray}
The first term of this expression is identically zero in case
of Gaussian initial conditions. 
The dominant contribution to the skewness is thus
given by the next-to-leading term, 
$3\mg\left(\kappa^{(1)}\right)^2\kappa^{(2)}\md$, that combines both
the expression of the linear amplification, and its second order expression.

To compute the previous expression it is worthwhile to know that
(Bernardeau 1995)
\begin{eqnarray}
&\displaystyle{\int\d^2\vk_1\int\d^2\vk_2\,W_{2D}\left(\vert\vk_1+\vk_2\vert\right)\,
W_{2D}(k_1)\,W_{2D}(k_2)}\times\nonumber\\
&\displaystyle{\ P(k_1)\ P(k_2)\ \left({5\over7}+{\vk_1\cdot \vk_2\over k_1^2}+
{2\over7}{(\vk_1\cdot \vk_2)^2\over k_1^2 k_2^2}\right)}\ =\label{sprop}\\
&\displaystyle{\int\d^2\vk_1\int\d^2\vk_2\ P(k_1)\ P(k_2)}\times\nonumber\\
&\displaystyle{\left({6\over 7}W_{2D}^2(k_1)
W_{2D}^2(k_2)+{1\over 4}W^2_{2D}(k_2){\d W^2_{2D}\over \d k}
(k_1)\right)}.\nonumber
\end{eqnarray}
Using this property we get,
\begin{eqnarray}
&\displaystyle{\mg\kappa^3(\theta_0)\md={6\over (2 \pi)^2}
\int_0^{\infty}{\d z'\over E(z')}\ w(z')^3\ {D_+^4(z')}}\times
\nonumber\\
&\displaystyle{\left[{6\over7}\left[\int k\d k W_{2D}^2(k)\,P(k)\right]^2+
\right.}\\
&\displaystyle{\left.{1\over 2}
\int k\,\d k\,W^2_{2D}(k)\,P(k)\times\int k\,\d k\,
W_{2D}(k)\,W_{2D}'(k)\,P(k)\right]}.\nonumber
\end{eqnarray}
This integral can thus be integrated numerically without more technical 
difficulties than for the second moment\footnote{Note that the property 
(\ref{sprop}) is valid for an angular top-hat window function only, and would
not be valid for a Gaussian window function for instance. 
For other types of window function
the dependence of the skewness on the shape of the power
spectrum is therefore expected to be more complicated.}.

At this stage it is crucial to notice that the ratio,
\begin{equation}
s_3(\theta_0)={\mg\kappa^3(\theta_0)\md\over \mg\kappa^2(\theta_0)\md^2},
\end{equation}
is  expected to be 
{\em independent of the normalization of the power spectrum}.
In the following we will explore in more detail the dependence of this ratio
on the different cosmological parameters and physical hypothesis.

To start with let us consider a very simple case in which we assume
that we have a power law spectrum, that all sources are at the same redshift
$z_s$ and that we live in an Einstein-de Sitter Universe. In such a case
we find that 
\begin{eqnarray}
\displaystyle{s_3(z_s)}&=\displaystyle{-\left[{36\over 7}-
{3\over 2}(n+2)\right]\,(n-1)\,(n-2)\,(n-3)^2}
\times\nonumber\\
&\displaystyle{\left[40-{4\,n\,(2\,n-1)\over 1+z_s}-{32\,n\over\sqrt{1+z_s}}
\right]/  }\label{s3z}\\
&\displaystyle{\left[16\,n\,(2\,n-5)\,(2\,n-3)\,(2\,n-1)\,
\left(2-2/\sqrt{1+z_s}\right)^2\right]}\nonumber\\
&\displaystyle{=-{5535\left(40+\displaystyle{{24\over 1+z_s}}+
\displaystyle{{48\over\sqrt{1+z_s}}}\right)
\over 32768\left(2-\displaystyle{{2\over \sqrt{1+z_s}}}\right)^2}\ \ 
{\rm for}\ n=-3/2};\nonumber
\end{eqnarray}
which gives,
\begin{equation}
\displaystyle{s_3(z_s)}=
\displaystyle{\approx -42\ \ {\rm for}\ n=-3/2\ \ {\rm and}\ \ z_s=1.}\nonumber
\end{equation}
This is  a quantity a priori
accessible to a measurement in a reasonably large catalogue. 
In the last section we will explore in more detail the expected
magnitude of the errors for doing such a measurement. 
Let us start with a study of the dependence of this quantity 
on the cosmological parameters.

\subsection{Numerical Results for a Realistic Case}


\begin{figure}
\vspace{5 cm}
\special{hscale=80 vscale=80 voffset=-50 hoffset=10 psfile=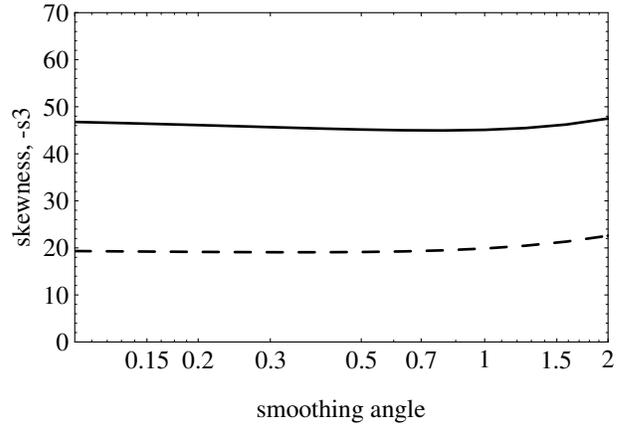}
\caption{The expected parameter $s_3$ as a function
of the smoothing angle (in degrees) for $z_s=1$ (solid line)  and $z_s=2$ (dashed
line)
in (\ref{nz}), the power spectrum (\ref{pk}) 
and assuming an Einstein-de Sitter Universe.}
\end{figure}

We present in Fig. 6 the results obtained for the function $s_3(\theta_0)$
for the power spectrum 
(\ref{pk}), the source distribution (\ref{nz}) with $z_s=1$ 
(solid line) and $z_s$=2 (dashed line)
in case of an Einstein-de Sitter Universe. We can see that the
function $s_3$ is rather flat, with little variation with the smoothing
angle.

\subsection{Dependence on the redshift}

\begin{figure}
\vspace{5 cm}
\special{hscale=80 vscale=80 voffset=-50 hoffset=0 psfile=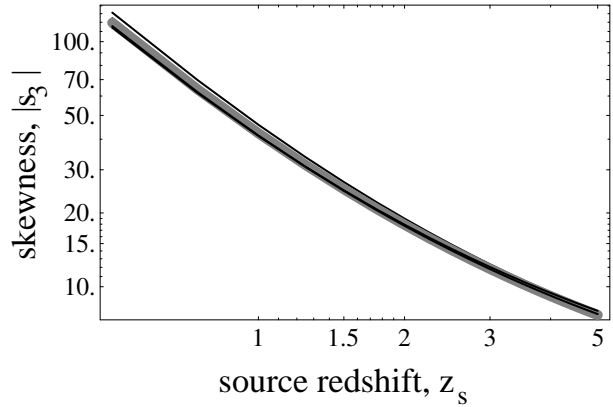}
\caption{The expected dependence of $s_3$ on the redshift
of the sources for an Einstein-de Sitter 
Universe. The curves have been plotted for different
power spectrum indices, $n=-1.3,\ -1.5,\ -1.9$
with the thick grey line for the intermediate value}
\end{figure}

From equation (\ref{s3z}) we easily can visualize the $z_s$ 
dependence of $s_3$.
This dependence is shown in Fig. 7. We can see that the result is almost
independent of the index $n$, 
and we have approximately,
\begin{equation}
s_3(z_s)\approx -42\ z_s^{-1.35}.
\end{equation}
As for the variance, we found that the dependence of the skewness on 
the redshift of the sources is also quite large.

%

\subsection{Dependence on the Cosmological Parameters}

\begin{figure}
\vspace{6 cm}
\special{hscale=80 vscale=80 voffset=-40 hoffset=10 psfile=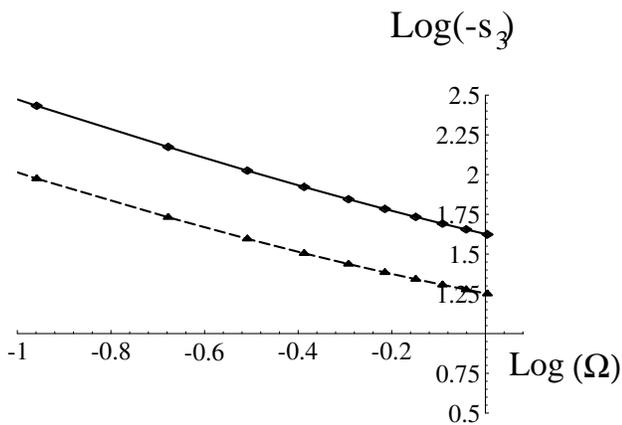}
\caption{The expected $\Omega_0$ dependence of $s_3$ for $z_s=1$
(solid line) and $z_s=2$ (dashed line).}
\end{figure}

\begin{figure*}
\vspace{7 cm}
\special{hscale=70 vscale=70 voffset=-10 hoffset=20 psfile=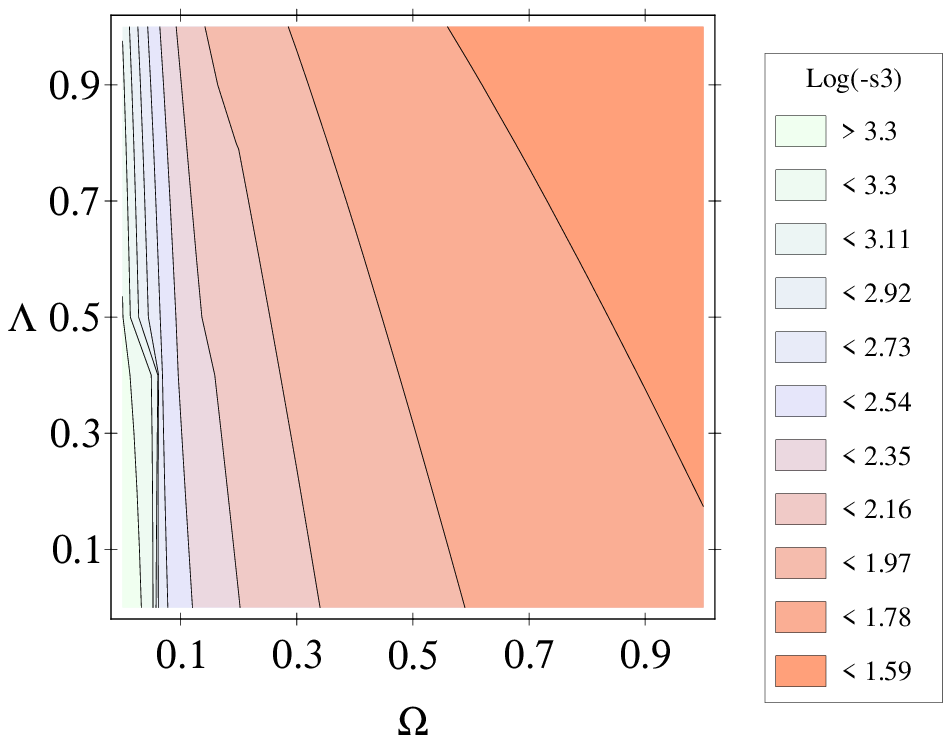}
\special{hscale=70 vscale=70 voffset=-10 hoffset=250 psfile=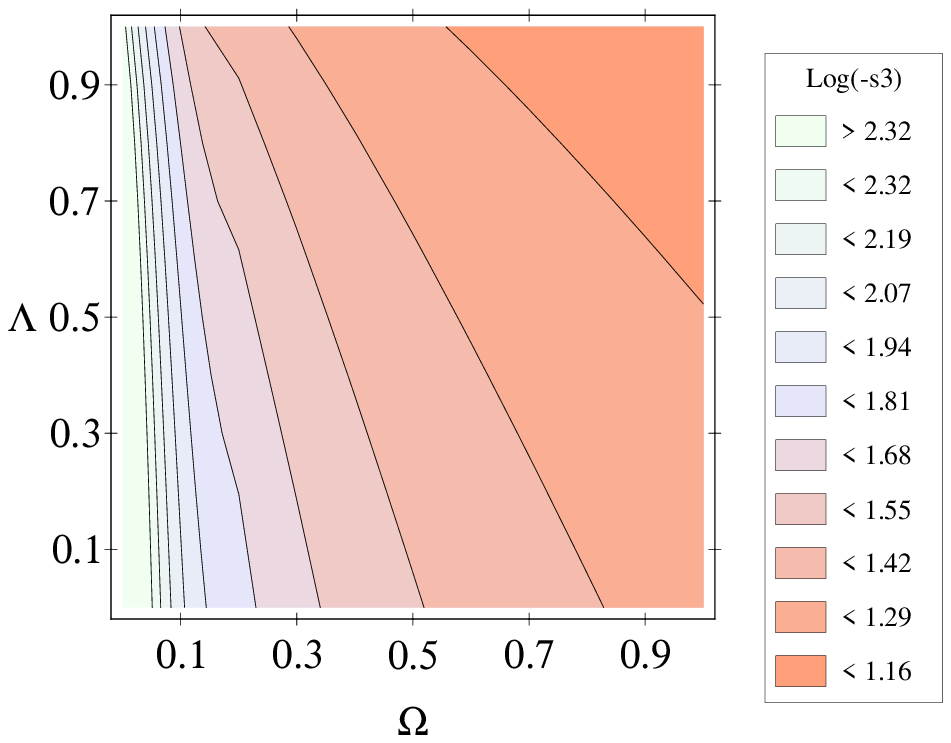}
\caption{Contour plot showing the dependence of $s_3$ on
$\Omega_0$ (horizontal axis)
and $\Lambda$ (vertical axis)
for $z_s=1$ (left panel) and $z_s=2$ (right panel). The contour lines
are the logarithmic values of $s_3$ regularly spaced in a logarithmic scale.}
\end{figure*}

From the general formulation we can also explore the dependence
of the results on the cosmological parameters.
In Fig. 8 we present the $\Omega_0$ dependence of the result
for $\Lambda=0$. We can see that the dependence is rather large,
approximatively as 
\begin{equation}
s_3(\Omega_0)\approx -42\ \Omega_0^{-0.8} 
\end{equation}
for $z_s \approx 1$.
 The $\Omega_0$ dependence is actually slightly weaker when the 
sources are at higher redshift (when the sources are at 
very small $z$ we recover the natural expectation that 
$s_3$ varies inversely with $\Omega_0$.)

In Fig. 9 we then present the contour plot showing the dependence
of the parameter on  $\Omega_0$ and $\Lambda$. We can see that
$s_3$ is significantly less $\Lambda$ dependent.

\subsection{Discussion, Simplified Model}

A this stage it is worth to compare the qualitative results
obtained here with those obtained for an a priori similar case: the
statistics in an angular survey. In both cases indeed the observed quantities 
are proportional to the cosmic density integrated along the line-of-sight.
The expected properties of the moments of the one-point
PDF are however qualitatively different. The reason is that
for a 2D angular survey, the density fluctuations are calculated
with a selection function implicitly normalized to unity, whereas
the efficiency function in the case of the weak lensing
is not normalized (see fig. 1). As a consequence the local convergence
is the {\sl summation} of the effects of each lens plane, whereas
the local measured galaxy over-density identifies with the
{\sl average} of all contributions present on the line-of-sight. The behavior
of these quantities is thus dramatically different when the
depth of the respective catalogues is changed.
The variance of the local convergence increases with the depth
(as for a random walk, the mean distance from the center increases
as the square root of the number of steps), but it becomes
more and more Gaussian, so that $s_3$ decreases. On the other
hand the variance of the local galaxy density contrast decreases
with the depth, since the density fluctuations tend to be smoothed out. 
The parameter $s_3$ is on the other hand almost independent 
of this depth. This is exactly what one would obtain by
considering respectively the summation or the average of a given
number of quasi-Gaussian variables. 

From these general remarks one can build a very simple model
for the moments of the local convergence that reproduces
qualitatively the results discussed in the previous sections.
Let us define $\wb(z_s)$ as the average of the efficiency function
when the sources are at redshift $z_s$,
\begin{equation}
\wb(z_s)=\int_0^{z_s}w(z){\d z\over E(z)}.
\end{equation}
Then, from eqs. (48) and (67) we simply expect that,
\begin{eqnarray}
\mg\kappa^2\md\propto&\displaystyle{{\wb(z_s)^2\over \mD_0^{n+3}(z_s)}}\\
s_3\propto&\displaystyle{-1\over \wb(z_s)},
\end{eqnarray}
which show the dependence of these statistical quantities on
the efficiency function and the depth of the catalogue.
These relations are certainly crude approximations, but
explain most of the features encountered previously, such
as the dependence of the results on $\Omega$
and on the redshift of the sources. 
It also suggests the existence of a combination of
$\mg\kappa^2\md$ and $s_3$ that would be almost independent
of the redshift of the sources. This combination (which is
almost the product of these two quantities, since $\wb(z_s)$ is 
proportional to the distance) would contain
information of purely cosmological nature.

\subsection{Systematics due to other Quadratic Couplings}

As it can be seen from Eq. (\ref{skew}) the skewness is induced by any
quadratic coupling in the observed convergence.  In particular
it implies that the convergence should not be measured
from the local shear using the linear approximation.
The method proposed by Kaiser (1995), or a similar method, that
takes into account the nonlinear relation between the local shear
and the local convergence has to be used.

Spurious couplings can appear also through:
\begin{itemize}
\item
nonlinear coupling between two deflecting lenses;
\item
coupling between two deflecting lenses due to the
induced displacement of the light path;
\item 
coupling between the population of selected sources and the local
convergence;
\item
density fluctuations of sources.
\end{itemize}

In the following we subsequently examine the importance of these different
effects.

\subsubsection{Coupling between two Deflecting Lenses}

This coupling is due to the fact that when 
the light path intercepts two lenses the resulting effect is 
given by  the product of the amplification
matrices. The resulting convergence $\kappa_{\rm tot}$
is given by the trace of this product, but coincides with the sum
of the convergences induced by each lens separately only 
in the linear regime. In general it is indeed given by
\begin{equation}
\kappa_{\rm tot}=\kappa_A+\kappa_B-\kappa_A\kappa_B-
\vec\gamma_A\cdot\vec\gamma_B
\end{equation}
which contain quadratic terms in the local lens densities
(the indices $A$ and $B$ correspond to quantities associated with
the lenses $A$ and $B$ respectively).

Actually this  nonlinear coupling between two lenses 
appears in the neglected term in (\ref{Kap2}).
This term induces a corrective expression for the
second order convergence term,
\begin{eqnarray}
&\displaystyle{\kappa^{(2)}_{\rm corr.(1)}(\gam)=
{3\over 4}\Omega_0\int_0^{\infty}{\d z'\over E(z')}\,w(z')}
\times\label{kapc1}\\
&\displaystyle{\int_0^{z'}{\d z''\over E(z'')}
{\mD_0(z'')\,\mD_0(z'',z')\over\mD_0(z')}\,
\phi^{(1)}_{ik}(\gam,z')\,\phi_{ki}^{(1)}(\gam,z'')}. \nonumber
\end{eqnarray}

The other source of coupling is due to the failure
of the Born approximation (i.e. the densities are calculated 
along the unperturbed light path) when two lenses are involved.
Indeed the convergence induced by a given lens
is not due to the local projected density at the observed
angular position if a foreground lens shifts the apparent
position of the background galaxies.
It implies that in (\ref{kappa1}) the local over-density
can be taken at the unperturbed position only at the
linear order but in general should be taken at the
position $\gam+\delta\gam$ where $\delta\gam$ is the displacement
induced by the foreground lenses. To get the
quadratic coupling one can simply write,
\begin{equation}
\delta^{(1)}_{\rm mass}(\gam+\delta\gam,z)\approx
\delta^{(1)}_{\rm mass}(\gam,z)+\nabla\delta^{(1)}(\gam,z)\cdot
\delta\gam.
\end{equation}
This term induces a term similar to (\ref{kapc1}),
\begin{eqnarray}
&\displaystyle{\kappa^{(2)}_{\rm corr.(2)}(\gam)=
{3\over 4}\Omega_0\int_0^{\infty}{\d z'\over E(z')}\,w(z')}
\times\label{kapc2}\\
&\displaystyle{\int_0^{z'}{\d z''\over E(z'')}
{\mD_0(z'')\,\mD_0(z'',z')\over\mD_0(z')}\,
\phi^{(1)}_{iik}(\gam,z')\,\phi_{k}^{(1)}(\gam,z'')}. \nonumber
\end{eqnarray}

These two terms obviously induce a corrective term for the 
skewness.
It can be easily calculated, and we estimate it here for an Einstein-de Sitter
universe for sources at a fixed redshift $z_s$. In such a case
the second order corrective term for the convergence reads,
\begin{eqnarray}
&\displaystyle{\kappa^{(2)}_{\rm corr.}(\gam)=
{9\over 8}
\int_0^{\mD_0(z_s)}\d\mD' {\mD_0(z_s)-\mD'\over \mD_0(z_s)}
\int_0^{\mD'}\d\mD''\,\mD''} \,\times\nonumber\\
&(\mD'-\mD'')\left[
\phi^{(1)}_{ik}(\mD')\,\phi_{ki}^{(1)}(\mD'')+
\phi^{(1)}_{iik}(\mD')\,\phi_{k}^{(1)}(\mD'')\right].
\end{eqnarray}
The resulting corrective term for $s_3$ is
\begin{equation}
s_3^{\rm corr.}=3\mg\kappa_{\rm corr.}^{(2)}
\left(\kappa^{(1)}\right)^2\md/\mg\left(\kappa^{(1)}\right)^2\md^2,
\end{equation}
so that\footnote{we use here a similar property as in (\ref{sprop}),
see Bernardeau (1995).}
\begin{eqnarray}
&\displaystyle{s_3^{\rm corr.}\approx 
\int_0^{\mD_0(z_s)}\d\mD'\int_0^{\mD'}\d\mD''}
\left[6-{3\over2}(n+2){\mD''\over\mD'}\right]\times\nonumber\\
&(\mD_0(z_s)-\mD')^2\,
(\mD_0(z_s)-\mD'')\,(\mD'-\mD'')\,(\mD'')^2\times\nonumber\\
&\mD'\,\mD_0(z_s)\,(\theta_0\,\mD')^{-2-n}
(\theta_0\,\mD'')^{-2-n}
/\nonumber\\
&\displaystyle{\left[
\int_0^{\mD_0(z_s)}\d\mD'
\left(\mD'\right)^2\,(\mD_0(z_s)-\mD')^2\,(\theta_0\,\mD')^{-2-n}\right]^2};\\
&\approx\ 1.55\ \ {\rm for}\ \ n=-3/2,
\end{eqnarray}
independently of the redshift of the sources.
It is completely negligible compared to the main contribution (\ref{s3z}),
and note that this effect is expected, at first glance, 
to be mainly independent of $\Omega_0$. 

It might be also worth to note that the resulting deformation
field induced by two lenses is no more potential (the 
amplification matrix is not symmetric). As a result,
the method proposed by Kaiser (1995) to build the local convergence
from the local shear is expected to fail 
at this level.
Once again, this is not expected to be crucial since the effects
of this coupling are seen to be small.

\subsubsection{Coupling between the Population of Sources and the Convergence}

The origin of this effect is the fact that the
population of sources on which  the shear is measured may change
with the local magnification.
In this case
\begin{eqnarray}
&\displaystyle{\kappa^{(2)}_{\rm corr.}=
-\kappa^{(1)}(\gam)\,
\int_0^{\infty}{\d z'\over E(z')}\,{\partial w(z')\over \partial\kappa}
\,\delta_{\rm mass}^{(1)}(\gam,z')},\label{kap2c3}
\end{eqnarray}
with
\begin{eqnarray}
{\partial w(z')\over\partial \kappa}=&\displaystyle{{3\over2}
\,\Omega_0\,(1+z')\,\mD_0(z')\,
\int_{z'}^{\infty}{\d z\over D_0(z)}\,{\partial
n(z)\over \partial \kappa}}\times\\
&\displaystyle{{1\over\sqrt{1\!-\!\Omega_0\!-\!\Lambda}}
\sinh\left[\sqrt{1\!-\!\Omega_0\!-\!\Lambda}
\int_{z'}^z{\d z''\over E(z'')}\right]}.\nonumber
\end{eqnarray}
where ${\partial n(z)/ \partial \kappa}$ is the partial 
derivative of the number density of sources at a given
redshift with the local
convergence (therefore the local magnification) for
a fixed normalization.
It can be evaluated a priori from the
population of objects used to do the measurements, and on the selection
procedure. For instance  the mean depth of the sources
is expected to change with the local magnification. 
We can build a simple model in which
we assume that the sources are always at a given redshift, 
but that its corresponding angular distance  is a function of the local
magnification,
\begin{equation}
\mD_0(z_s)=\mD^0\ (1+\alpha\,\kappa).
\end{equation}
It would imply, for an Einstein-de Sitter universe,
\begin{equation}
{\partial w(z)\over \partial \kappa}=
\alpha\ {3\over 2}{\mD_0^2(z)\over \mD^0}.
\end{equation}
In such a case the corrective term for $s_3$ can be calculated
straightforwardly and is given by,
\begin{eqnarray}
s_3^{\rm corr.}=&9\,\alpha\,(1-n)/2,\\
\approx&11\,\alpha\ \ {\rm  for}\ \ n=-1.5.
\end{eqnarray}
It is not necessarily completely negligible, depending on the value
of $\alpha$, that is the sensitivity of the selected population
of sources with the local magnification (this will be explored in more
details in the second paper).
So, one way to reduce the effect of this coupling 
is to select the galaxies used to measure the shear 
from a surface brightness criterion (Bartelmann \& Narayan 1995).
In practice however, it is probably
impossible to suppress all influence in the selection procedure
of the total luminosity of the objects.

\subsubsection{Coupling due to Density fluctuations of Sources}

Throughout this study we have always implicitly assumed that the
sources formed a uniform background population.
This is actually far from being true, and one naturally 
expects the sources to have cosmic density fluctuations.
The effects of these fluctuations of the 
measurements of the local convergences strongly depend  on the
adopted method. For a method based on the estimation of the
local magnification by the depletion of the galaxy number density
 (magnification bias), this can be a very strong
effect on both moments.
For methods in which the local convergence is built from
the measured shear, this effect is weak.
Simple considerations show that it introduces a significant
coupling only at the level of the fourth moment.
This is due to the fact that the populations of sources
and lenses can be considered as two independent 
populations so there are no direct couplings between their
density fluctuations.

\section{Error Evaluations due to the Cosmic Variance}

In this section we investigate the effects of the cosmic variance on the
precision of the measurements of the second moment and on $s_3$.
The origin of this error is the finite size of the sample in which
the measurement could be done. Typically we will assume that the
sample size is about $25\ deg^2$, which could be 
accessible in the coming years for the now available technologies.
In this section we describe the measured local convergence,
$\kappa_{\rm meas.}(\gam)$ as the sum of the exact local
convergence and a noise $\epsilon(\gam)$,
\begin{equation}
\kappa_{\rm meas.}(\gam_i)=\kappa(\gam_i)+\epsilon(\gam_i).
\end{equation}
The random noise $\epsilon(\gam_i)$ components are assumed to be
all independent of each other, independent of the true
values of the local convergence and to have the same variance
$\epsilon$,
\begin{equation}
\mg\epsilon^2(\gam_i)\md^{1/2}=\epsilon.
\end{equation} 
It is reasonable to assume that $\epsilon$ is of the order
of $10^{-3}$ (see Appendix B).
We then introduce the notation $\Xb$ which is the
(connected) geometrical average of the quantity $X$ in the sample.
These quantities are also cosmic random variables and are thus expected
to vary from one sample to another. Our goal is then to estimate the
ensemble averages and variances of those geometrical averages
for the quantities of interest, i.e. to know how good they
are as estimates of the true cosmic quantities.

 To do such calculations we have to introduce the expression of
joint moments of the local convergence. 
Luckily, this is not too complicated!
Indeed in the quasi-linear regime one can show that, when
the separation between two cells is large compared to the smoothing angle,
we have 
\begin{eqnarray}
&\mg\kappa^p(\theta_0,\gam_1)\ \kappa^q(\theta_0,\gam_2)\md=\nonumber\\
&c_{pq}\ \mg\kappa(\theta_0,\gam_1)\ \kappa(\theta_0,\gam_2)\md\ 
\mg\kappa^2(\theta_0)\md^{p+q-2},
\end{eqnarray}
where $c_{pq}$ are finite quantities.
This is the application to 2D statistics 
of what was developed in 3D by Bernardeau (1996a). However
contrary to the 3D case, and due to projection effects, these
coefficients depend on the details of the adopted model.
They can anyway be calculated numerically (see Appendix A).
Using these results we can calculate both the expectation value of a
measured quantity and the variance of this measure. Thus we have 
\begin{eqnarray}
&\displaystyle{{\mg\ovl{\kappa_{\rm meas.}^2(\theta_0)}\md-\mg\kappa^2(\theta_0)\md
\over\mg\kappa^2(\theta_0)\md}=
{\epsilon^2-\mg \kappa^2(\theta_{\rm sample})\md 
\over \mg \kappa^2(\theta_0)\md}};\\
&\displaystyle{{\Delta\mg\kappa_{\rm meas.}^2(\theta_0)\md\over 
\mg\kappa_{\rm meas.}^2(\theta_0)\md}=
\sqrt{c_{22}\ \mg\kappa^2(\theta_{\rm sample})\md}};\\
&\displaystyle{{\mg s_3^{\rm meas.}\md-s_3 \over s_3}}=\nonumber\\
&\displaystyle{\left(2-3{c_{21}\over s_3}\right)\ 
{\mg\kappa^2(\theta_{\rm sample})\md \over \mg\kappa^2(\theta_0)\md}-
2\ {\epsilon^2\over\mg\kappa^2(\theta_0)\md}};\\
&\displaystyle{{\Delta s_3^{\rm meas.}\over s_3^{\rm meas.}}}=\nonumber\\
&\displaystyle{\sqrt{ \left(
{c_{33}\over s_3^2}+4 c_{22}-4{c_{32}\over s_3}\right)\ 
\mg\kappa^2(\theta_{\rm sample})\md}}.
\end{eqnarray}
The numerical calculations give the values of the required coefficients 
$c_{pq}$ (Appendix A), from which we have 
\begin{eqnarray}
&\displaystyle{{\mg\ovl{\kappa_{\rm meas.}^2(\theta_0)}\md-\mg\kappa^2(\theta_0)\md
\over\mg\kappa^2(\theta_0)\md}=-0.13};\\
&\displaystyle{{\Delta\overline{\kappa_{\rm meas.}^2(\theta_0)}\over 
\mg\kappa_{\rm meas.}^2(\theta_0)\md}=0.14};\\
&\displaystyle{{\mg s_3^{\rm meas.}-s_3\md \over s_3}=-0.07};\\
&\displaystyle{{\Delta s_3^{\rm meas.}\over s_3^{\rm meas.}}=0.03.}
\end{eqnarray}
The cosmic errors of course decrease with the size of the sample.
What the previous results show is that the errors are 
already quite small when the size of the sample is $25\ deg^2$ 
and would allow, for a perfectly well known source distribution,
a determination of the cosmological parameters at the 10 to 15\%
level.

\section{Discussion and conclusions}

Using Perturbation theory techniques, we have calculated
the variance and the skewness of the local
filtered convergence $\kappa$ in an
inhomogeneous cosmology in the FRW framework. The quantity $\kappa$
can be calculated from the measurement of the magnification $\mu$ and/or the
distortion $\delta$.

The exploration of the dependence of the variance of  
$\kappa$ on the parameters of interest have shown that
it is, as expected, strongly dependent on $\Omega_0$,
but also on $\Lambda$. It is of course
proportional to the magnitude of the power spectrum
of the matter fluctuations. These dependences, although degenerate
with each other, are of immediate cosmological interest.
Note that these dependences are given for a given power spectrum
normalization at redshift zero. As pointed out in other
studies (Kaiser 1996, Jain \& Seljak in preparation) if the 
power spectrum is normalized to the present number density of clusters
the degeneracy with $\Omega$ almost vanishes.  
More worrying is the fact that
the variance is also strongly dependent on the redshift of the
sources $z_s$, making difficult to analyze quantitative
results in a cosmological context if the sources are not well known.

It is possible to disentangle
the dependence of the variance on the magnitude of the power spectrum
and the cosmological parameters by the use
of two different populations of sources.
Thus, the ratio of the variances for these two populations
could provide constraints on the cosmological parameters 
independently of the power spectrum (the dependence on its shape
is only weak). Of course the problem mentioned previously
for the knowledge of the population of sources is
even more critical since both populations have to be perfectly known.

More interestingly, the skewness  $s_3$ of the
local convergence, expressed as the ratio of the third moment
and the square of the second, can also provide a way
to separate the dependence on the cosmological parameters.
We indeed found that this quantity does not  depend
 on the amplitude of the power spectrum, is weakly
dependent on its shape  and varies almost as the inverse of the
density parameter, with a slight degeneracy with the
cosmological constant.
Moreover the skewness is a probe of the initial Gaussian nature 
of the density field. If physical mechanisms such as cosmic strings,
topological defects, symmetry breaking, etc... generate non Gaussian features 
in the initial density field, they will modify the behavior of $s_3$
(see Gazta\~naga \& M\"ah\"onen 1996), 
inducing a strong angular variation of $s_3$.
The dependence of $s_3$ 
on the redshift of the sources is surprisingly large, 
stressing the absolute necessity
of knowing the population of sources used to do such measurements.
It is however interesting to note that
the product of the variance and the skewness is almost independent
of the redshift of sources, so in any case we have a robust estimator
of the product $P(k) \ \Omega^{0.5}$. It might reveal to be an interesting 
quantity to consider. 

%

An important concern was the contribution of multiple lenses configurations 
in  reducing the significance of these statistical estimators.  We have
shown that the two main quadratic couplings between two lenses  are   
ineffective, with contributions of a few percent, and do not strongly
depend on the cosmology. Thus, whatever the cosmological model, the 
cosmological signal contained in the variance and the skewness could be
slightly weakened, but  is not  washed out by these additional couplings.

We have not considered higher order moments, basically because we think
that they are not going to improve the situation, i.e. to raise
the degeneracy between the cosmological parameters
we have with the second and third moments. 
A rough calculation (taking into account only the dominant
contribution) shows interestingly  that the
$s_4$ to $s_3^2$ ratio is expected to be close to 2 for any 
cosmological model ($\Omega_0$, $\Lambda$ and the shape of the
power spectrum). In principle it could be used as a test for the 
gravitational instability scenario with Gaussian initial conditions.
We are however far from a reliable measurement of all these quantities.

A concern that has not been addressed in this paper is the validity
domain of these results with respect to the nonlinear corrections
due to the dynamical evolution of the cosmic fields. 
These corrections are expected to intervene at small angular scale where
the physical scale of the lenses is expected to enter the
nonlinear regime. The question is naturally, at which 
angular scale? The answer is not straightforward, and actually 
the situation is slightly different for the second moment and for the
skewness. 
Indeed the second moment is always a measure
of the projected power spectrum whether or not it is in the
linear regime. This point was clearly made in a recent work
(Jain and Seljak, in preparation) where the rms polarization is
calculated including the nonlinear evolution
of the power spectrum using prescriptions as the ones developed
by Hamilton et al. (1991), Jain, Mo \& White (1995) and
Peacock \& Dodds (1996). 
For $s_3$ there are no such transformations that would take
into account the nonlinear effects. But one should have in mind that 
this parameter has been observed to be very robust against
nonlinear corrections (this is not necessarily the case for the product
of the variance and $s_3$). It has been amply checked in numerical simulations
for the 3D filtered density field (see the remarkable results of
Baugh, Gazta\~naga \& Efstathiou, 1995). The situation
for the projected density field has not been investigated
in as many details, and extended numerical
work would be necessary for that respect. 
In particular it
would be interesting to extend the work of Colombi et al. (1996),
who propose a prescription to correct the high order
moments of the local 3D filtered density PDF, to the
projected density. Another arduous line of investigation
would be to take into account in the analytic
calculations the so called ``loop corrections''
as it has been pioneered by Scoccimarro \& Frieman (1996)
for the un-smoothed density field. Such calculations 
could quantify the errors introduced by the nonlinear effects
at small angular scale.

In the last section we have briefly explored the 
observational requirements  for completing a large observational program. 
We have in particular tried to estimate
the cosmic noise that would affect a catalogue covering an area 
of 25$\ deg^2$. This calculation is at most indicative
and should not be considered too realistic.
For instance, in this analysis we have made a fairly crude hypothesis
for the noise.
In particular we have assumed that the noise due to the intrinsic
ellipticities of the galaxies is local. If this is true for the 
local distortion measurement, this is not true for the convergence
if it has been obtained from the reconstruction scheme proposed
by Kaiser since it involves the gradient of $\kappa$.
In this analysis we have also not taken into account
the fluctuations in the number density of sources, that
could be significant at the angular scale we are interested in.

On the other hand we have assumed that the moments were calculated
directly, i.e. with the average of the proper power
of the different measured convergences. This may not be the most robust way 
to do it. Indeed, taking advantage of the fact that we expect the
Probability Distribution Function of the local convergence
to be close to a Gaussian distribution it is
possible to estimate the variance and the skewness simply from the
shape of the PDF around its maximum. More precisely 
one can use the Edgeworth expansion (see Juszkiewicz et al. 1995
and Bernardeau \& Kofman 1995) to look for the best
fitting values of the low order moments without having to
actually compute them. With such a method the results
should be less sensitive to the cosmic noise, because less
sensitive to the rare events in the tails.

We leave for a coming paper the difficult task of
defining what we think could be an optimal 
strategy for observing the shear induced by the large-scale structures 
and obtaining constraints of cosmological interest out of it.
But we stress in any case that such 
a project demands a lensing survey at 25 square degrees
scale with a very high image quality. 
Such a lensing survey will be used for the statistical work described in
this paper, but also to probe the angular power spectrum at scales from
the Mpc to few hundred Mpcs.
This program is accessible with
the UH8K camera or with the future 16K$\times$16K (MEGACAM) camera at the 
Canada-France-Hawaii Telescope, provided optimal algorithms for
measuring very weak shear are used. The pixel to pixel autocorrelation
function (ACF) proposed by Van Waerbeke et al. (1996) which consists in 
analyzing the ellipticity of the ACF induced by the gravitational shear
seems a promising
approach because it does not require any more to define object centroids and to
compute shape parameters for each individual faint galaxy from noisy
images.
The shape of the sample also impacts on the signal to noise:
 reconstruction method {\sl \`a la Kaiser} is expected to introduce
finite size errors along the edges of the catalogue, so it would
be interesting to have a catalogue as compact as possible. On the other hand
an elongated catalogue has less cosmic variance.

The drawback of this project is that it requires the knowledge of the
redshift distribution of the sources, especially for faint objects
($B>25$). A spectroscopic survey to get the redshift of these objects is
unrealistic with present-day ground based telescopes; fortunately many other 
methods are possible, such as the
photometric redshifts, the lensing  inversion technique, and the
depletion method (see Mellier 1996 for a review). None of them have proved 
their ability to get
secure redshifts, but they are potentially very attractive because they are
not time consuming.


\section*{Acknowledgments}

F. B. is grateful to IAP 
where most of this work has been conducted
and to the Observatoire Midi-Pyr\'en\'ees where it has been completed. The authors
are also thankful to the Observatoire de Strasbourg for its 
very warm hospitality.

\section*{Appendix A: Expressions of joint cumulants}

\subsection*{General Expressions}

The statistical quantities described in the main text are expected
to be affected by a cosmic variance due to the finite size of the
sample. The estimation of such an effect is determined by joint moments
between local smoothed convergences taken at two different locations.
Indeed these moments provide
information on how observed local smoothed convergences are 
expected to be correlated. The exact derivation of the joint moments 
of interest would be a difficult task, that requires
an explicit calculation of the higher  order expression (up to
the sixth order!) of
the local convergence. As we do not require a very accurate
calculation but rather a realistic estimation of these quantities
we will make some simplifications in this derivation.
First of all we will assume that the high order terms of the
local convergence are dominated  by the projection along
the line-of-sight of the corresponding high order terms of the
local density in the dominant lens plane. This is what has been
found in Section 5 for the skewness, and we expect it to be also true
for the cumulants we are interested in.
Moreover to do the calculations we will assume that the size of the sample
in which the local convergence is available is significantly
larger that the smoothing angular scale. This will make the derivation of the
joint cumulants more simple and accessible to Perturbation
Theory. Similar results were 
obtained in a slight different context (Bernardeau 1996a) for which
the derivation of joint cumulants is given for 3D top-hat 
smoothed density contrasts.
It is shown that when the distance between the smoothing cells
is large compared to the smoothing scale, the joint cumulants
can be written,
\begin{equation}
\mg\delta_1^p\,\delta_2^q\md_c\approx\,C_{p,q}\,\mg\delta_1^2\md^{p-1}
\,\mg\delta_2^2\md^{q-1}\,\mg\delta_1\,\delta_2\md,\label{jcul}
\end{equation}
at the leading order in Perturbation Theory, and where the $C_{pq}$
coefficients are finite numbers (they are constant for a power law 
spectrum). A crucial property for these coefficients is that
we have,
\begin{equation}
C_{p,q}=C_{p,1}\,C_{q,1}
\end{equation}
so that the derivation of these coefficients reduces to the
calculations of a single index series of numbers. They have
been given in the paper mentioned above for the 3D top-hat smoothing.

These results cannot be used directly to estimate the joint moments we are 
interested in, but form however the basis of the calculations.
Indeed if we assume, as mentioned in the introduction of this
appendix, that the high order expressions of the local convergence are
dominated by the projection of the local density of the lenses at the
same order, then the expressions of the joint moments are given by the
integration over the line-of-sight of the relation (\ref{jcul})
where the coefficients $C_{p,q}$ correspond to the top-hat smoothed
joint cumulants {\sl for the 2D dynamics}. This is a situation which is
similar to the situation encountered in the derivation of the 
expression of the high order moments in angular survey.

Using the calculation developed in Bernardeau (1996a) and applying
it to the 2D dynamics we get, for a power law spectrum of
index $n$,
\begin{equation}
C^{2D}_{2,1}={24\over 7}-{(n+2)\,\over 2},
\end{equation}
and 
\begin{equation}
C^{2D}_{3,1}={1473\over 49}-{195\over 14}\,(n+2)+{3\over 2}\,(n+2)^2.
\end{equation}

Then, as a consequence, the local joint cumulants follow a similar hierarchy
\begin{eqnarray}
&\mg\kappa(\gam_1)^p\,\kappa(\gam_2)^q\md_c=\label{convhie}\\
&c_{p,q}\,\mg\kappa(\gam_1)^2\md^{p-1}\,\mg\kappa(\gam_2)^2\md^{q-1}\,
\mg\kappa(\gam_1)\,\kappa(\gam_2)\md,\nonumber
\end{eqnarray}
where the coefficients $c_{p,q}$ are given by ratios of integrals
along the line-of-sight.
We give here the expressions of these integrals for an Einstein-de Sitter
Universe, and we will limit the explicit numerical calculations of these
expressions for this case,
\begin{equation}
c_{p,q}=C^{2D}_{p,q}\,{Q_{p+q}\over Q_{2}\,P_2^{p+q-2}}.
\end{equation}
with
\begin{eqnarray}
&\displaystyle{Q_p=\theta_0^{-(n+2)(p-2)}}\\
&\displaystyle{\int_0^{\mD_0}\d \mD\,w(\mD)^p\,
\mD^{(n+2)(p-2)}\, D_+^{2p-2}(\mD)\,\xi_{2D}(\mD)}\nonumber\\
&\displaystyle{P_p=\theta_0^{-(n+2)(p-1)}}\\
&\displaystyle{\int_0^{\mD_0}\d \mD\,w(\mD)^p\,
\mD^{(n+2)(p-1)}\, D_+^{2p-2}(\mD)}\nonumber
\end{eqnarray}
where
\begin{equation}
\xi_{2D}(\mD)=\int_0^{\infty}
\kort\ \d \kort\ P(\kort)\ J_0[\kort\,\mD\,\theta_{1,2}].
\end{equation}
$J_0$ is the Bessel function of the first kind and
$\theta_{1,2}$ is the 
angle between the two directions of the smoothing cells 1 and 2.

The final step is to take the geometrical averages of these
joint moments in the sample. It is expected to be dominated  by cells
being at the distance of the order of the size of the
sample, $\theta_{\rm sample}$. Then the averaged moments follow the
same hierarchy as in (\ref{convhie}) but for which the function $\xi(\mD)$
is replaced by 
\begin{equation}
\xib_{2D}(\mD)=\int_0^{\infty}
\kort\ \d \kort\ P(\kort)\ W_{2D}^2[\kort\,\mD\,\theta_{1,2}].
\end{equation}
where $W_{2D}$ is introduced in (\ref{w2d}). This result
is valid if the sample has a spherical shape, although it is by no means 
a crucial hypothesis.

In the latter case the integral $Q_2$ is actually the expression
of the variance of the convergence in the  whole sample. All the 
considered joint cumulants are also found to be
proportional to this variance, and thus the cosmic 
uncertainties on the results 
will be all the more important that this integral
is large. The results will therefore depend very strongly on the
adopted shape of the power spectrum. In the following subsection we give the
numerical results for the APM measured power spectrum (\ref{pk}).
Compared to a CDM spectrum it contains more power at large
scale and is thus less favorable for the estimation  of the
cosmic errors, but it is probably more realistic.

\subsection*{Numerical Results}

In the following we give the results of the previous integrals
for the power spectrum given in (\ref{pk}). The index at the 30' smoothing
scale is found to be
\begin{equation}
n\approx -1.45,
\end{equation}
and then the coefficients of interest are found to be
\begin{eqnarray}
c_{2,1}=&36.7\\
c_{2,2}=&1400\\
c_{3,1}=&3200\\
c_{3,2}=&125,300\\
c_{3,3}=&11,400,000
\end{eqnarray}
for a sample size of 25 $deg^2$, corresponding to a sample
size of $\theta_{\rm sample}\approx 3\ deg$.
These results are calculated with the distribution of lenses 
(\ref{nz}) in which $z_s\approx 1$.

\section*{Appendix B: Evaluation of the Cosmic Errors}

In this appendix we present the derivation of the errors,
due to the cosmic variance, i.e. the fact that measurements made in a finite
sample are affected by systematic and random errors. We will
naturally focus our presentation on the second and third moments 
of the local smoothed convergence.

To present the calculations it is necessary to introduce two different
averages. One is the geometrical average, the mean value of a given 
observed quantity at different locations, and the second is the 
ensemble average, which corresponds to the expectation value
of quantities that depend on  cosmic variables (local densities..)
or random quantities associated with intrinsic properties
of the galaxies such as their {\sl intrinsic} ellipticities.
In principle the two averages of a same quantity coincide
in a perfect case, that is if the available data set was infinite, 
but in practice
this is obviously not the case!
Actually, the geometrical averages, considered as estimations
of the corresponding ensemble averages, are 
also random variables that are expected to vary from one
sample to another. In the following we estimate
the properties of these geometrical averages, that is their 
cosmic expectation values and their variances, assuming that
for a sufficiently large sample they obey a Gaussian
statistics. We follow here ideas that were developed 
by Colombi, Bouchet \& Schaeffer (1995) and more precisely by
Szapudi \& Colombi (1996).

\subsection*{Notations}

In the following we denote $\Xb$ the (connected) geometrical average
of a quantity $X$  (and as usual, $\mg X\md_c$ its ensemble
average). Thus the observed moments of the measured convergence
are 
\begin{eqnarray}
\ovl\kappa&=\displaystyle{{1\over N_c}\sum_{i=1}^{N_c}\,\kappa_i}\\
\ovl{\kappa^2}^c&=\displaystyle{{1\over N_c}\sum_{i=1}^{N_c}\,\kappa_i^2-(\ovl\kappa)^2}\\
&\dots\nonumber
\end{eqnarray}
where the summations are made over the $N_c$ different locations where
the smoothed convergences  are supposed to have been determined. We actually
assume that the local convergences have been measured in a compact
sample, of a size significantly larger than the smoothing length.
In the geometrical averages it is also {\sl not} excluded 
that the smoothing areas overlap, so that $N_c$ can actually be arbitrarily 
large. Of course the resulting averages cannot be arbitrarily
accurate because the measured $\kappa_i$ are  correlated
variables, and will be all the more correlated that they are measured
in close directions. 

\subsection*{Modelisation}

In the following we  not only estimate the errors due to the
cosmic variance but also the effects of the intrinsic noise in the
measurements due to the use of a limited number of tracers
for the shear and of their intrinsic imperfections. 
So assume that the measured convergence in the direction $\gam_i$ 
is given by,
\begin{equation}
\kappa_{\rm meas.}(\gam_i)=\kappa(\gam_i)+\epsilon(\gam_i),
\end{equation}
where $\kappa(\gam_i)$ is the true cosmological value
of the local convergence in the direction $\gam_i$ and
$\epsilon(\gam_i)$ is the error made in this same direction
due to the intrinsic ellipticities of the galaxies.
Assuming the ellipticities of the galaxies are independent
from one another, and using results obtained
for their intrinsic ellipticities (see for instance Miralda-Escud\'e
1991b) we get,
\begin{equation}
\mg\epsilon(\gam_i)^2\md\approx{10^{-1}\over \sqrt{n_g}},
\end{equation}
where $n_g$ is the number of galaxies in a given smoothing 
angular cell.
Moreover two variables $\epsilon(\gam_i)$
and $\epsilon(\gam_j)$ are independent if the corresponding cells
do not overlap, and they are all assumed to be independent of the
true values of the local convergences. This is probably a rather
simple and naive modelisation but we leave more accurate numerical studies
of a more realistic modelisation  for a further paper.

As the variance of $\epsilon(\gam_i)$ is independent of $i$
we will denote, $\epsilon$ their common RMS fluctuation, 
\begin{equation}
\mg\epsilon(\gam_i)^2\md=\epsilon^2
\end{equation}
and from the expected density of galaxies in a $(30')^2$
field (about $10^4$ for the limit magnitudes $B=27$ or $I=26$ currently used 
in ultra-deep imaging; Tyson 1988, Smail et al. 1995)
we can estimate that
\begin{equation}
\epsilon\approx 10^{-3}.
\end{equation}

We then define the random variable $e_p$ that describes
the departure between an observed geometrical average 
and the corresponding ensemble average,
\begin{equation}
\ovl{\kappa(\theta_0)}=e_1
\end{equation}
and 
\begin{equation}
\ovl{\kappa^p(\theta_0)}^c=\mg{\kappa^p(\theta_0)}\md_c\ (1+e_p)
\end{equation}
for $p>1$.
In the following we will assume that the cosmic random
variables $e_p$ have a small
variance compared to the variance at the smoothing scale. Therefore
the variables $e_p$ can be considered to follow a Gaussian statistics
and we will do the calculation at their dominant order.
The knowledge of the cosmic errors in any of the considered measurements
can then be derived from the values of the variances and cross-correlations 
between these variables.  

\subsection*{Statistical properties of the variables $e_p$}

To obtain the statistical properties of $e_p$ one can simply consider
the ensemble average of the $\ovl{\kappa^p(\theta_0)}^c$, 
of their squares and products.
From the assumed properties of the variables $\epsilon(\gam_i)$
and $\kappa(\gam_i)$ we have,
\begin{eqnarray}
\mg\ovl{\kappa_{\rm meas.}(\theta_0)}\md&=0\\
\mg\ovl{\kappa_{\rm meas.}(\theta_0)}^2\md&=
\mg \kappa^2(\theta_{\rm sample})\md
\end{eqnarray}
where we have identified the geometrical average of 
$\mg\kappa(\gam_i)\,\kappa(\gam_j)\md$
with the variance of the local convergence at the scale
of the sample, $\theta_{\rm sample}$. At this scale we assume that
the ensemble average of $\epsilon(\gam_i)$ is negligible compared
to the one of $\kappa$: it 
decreases like Poisson noise, that is more rapidly than the
variance of the convergence.

The ensemble averages for the second moment read,
\begin{eqnarray}
&\mg\ovl{\kappa^2_{\rm meas.}(\theta_0)}^c\md=\nonumber\\
&\mg \kappa^2(\theta_0)\md+\epsilon^2-
\mg \kappa^2(\theta_{\rm sample})\md \\
&\mg\left(\ovl{\kappa^2_{\rm meas.}(\theta_0)}^c\right)^2\md=\nonumber\\
&\mg \kappa^2(\theta_0)\md^2-2\ \mg \kappa^2(\theta_0)\md\ 
\mg \kappa^2(\theta_{\rm sample})\md+\\
&+2\ \epsilon^2\ \mg \kappa^2(\theta_0)\md+
c_{2,2}\ \mg \kappa^2(\theta_0)\md^2\ \mg \kappa^2(\theta_{\rm sample})\md
\nonumber
\end{eqnarray}
where the calculations have been limited to the quadratic
terms in $\epsilon$ and $\mg \kappa^2(\theta_{\rm sample})\md^{1/2}$.
The calculations make also intervene the geometrical
averages of the joint moments
$\mg \kappa^2(\gam_i)\,\kappa^2(\gam_j)\md$ which are evaluated 
in the previous appendix.

For the third moment we have,
\begin{eqnarray}
&\mg\ovl{\kappa^3_{\rm meas.}(\theta_0)}^c\md=\nonumber\\
&\mg \kappa^3(\theta_0)\md-3\ c_{2,1}
\mg \kappa^2(\theta_0)\md\ \mg \kappa^2(\theta_{\rm sample})\md, \\
&\mg\left(\ovl{\kappa^3_{\rm meas.}(\theta_0)}^c\right)^2\md=\nonumber\\
&\mg \kappa^3(\theta_0)\md^2+c_{3,3}
\mg \kappa^2(\theta_0)\md^4\ \mg \kappa^2(\theta_{\rm sample})\md -\\
&-6\ c_{2,1}\mg \kappa^2(\theta_0)\md^3\ 
\mg \kappa^2(\theta_{\rm sample})\md\ .\nonumber
\end{eqnarray}
We have also to consider the cross-correlations between those
moments which give,
\begin{eqnarray}
&\mg\ovl{\kappa_{\rm meas.}(\theta_0)}^c\ 
\ovl{\kappa^2_{\rm meas.}(\theta_0)}^c\md=\nonumber\\
&c_{2,1}\,\mg \kappa^2(\theta_0)\md\ \mg \kappa^2(\theta_{\rm sample})\md\,, \\
&\mg\ovl{\kappa_{\rm meas.}(\theta_0)}^c\ 
\ovl{\kappa^3_{\rm meas.}(\theta_0)}^c\md=\nonumber\\
&c_{3,1}\,
\mg \kappa^2(\theta_0)\md^2\ \mg \kappa^2(\theta_{\rm sample})\md\,, \\
&\mg\ovl{\kappa^2_{\rm meas.}(\theta_0)}^c\ 
\ovl{\kappa^3_{\rm meas.}(\theta_0)}^c\md=\nonumber\\
&\mg \kappa^3(\theta_0)\md^2-
\mg \kappa^3(\theta_0)\md\ \mg \kappa^2(\theta_{\rm sample})\md+\\
&+c_{3,2}\ \mg \kappa^2(\theta_0)\md^3\ \mg\kappa^2(\theta_{\rm sample})\md-
\nonumber\\
&-3\ c_{2,1}\ \mg \kappa^2(\theta_0)\md^2\ \mg\kappa^2(\theta_{\rm sample})\md+
\epsilon^2\ \mg \kappa^3(\theta_0)\md.\nonumber
\end{eqnarray}
From these results we have entirely define the statistical
properties of the variables $e_1$, $e_2$ and $e_3$.
Simple identifications lead to 
\begin{eqnarray} 
\mg e_1\md&=&0,\\ 
\mg e_1^2 \md&=&\mg \kappa^2(\theta_{\rm sample})\md,\\
\mg e_2\md&=&{\epsilon^2-\mg \kappa^2(\theta_{\rm sample})\md \over 
\mg \kappa^2(\theta_0)\md},\\ 
\mg e_2^2 \md&=&c_{2,2}\ \mg \kappa^2(\theta_{\rm sample})\md\\
\mg e_3\md&=&-3 {c_{2,1}\over s_3}\,
{\mg \kappa^2(\theta_{\rm sample})\md \over \mg \kappa^2(\theta_0)\md},\\ 
\mg e_3^2\md&=&{c_{3,3}\over s_3^2}\ \mg \kappa^2(\theta_{\rm sample}) \md,\\
\mg e_1 e_2 \md&=&c_{2,1}\ \mg \kappa^2(\theta_{\rm sample})\md,\\ 
\mg e_1 e_3 \md&=&{c_{3,1}\over s_3}\ \mg \kappa^2(\theta_{\rm sample})\md,\\ 
\mg e_2 e_3 \md&=&{c_{3,2}\over s_3}\ \mg \kappa^2(\theta_{\rm sample})\md.
\end{eqnarray}
It is interesting to note that $\epsilon$ enters only in the
expectation value of $e_2$. 

As a result we can calculate both the expectation value of a
measured quantity and the variance of this measure. 
To do so the quantities we are interested in are expressed
in terms of the random variables $e_p$ and expanded
up to the quadratic order. This is a trivial calculation for the
variance, and for the measured $s_3$ we have
\begin{eqnarray}
s_3^{\rm meas.}&=\displaystyle{s_3{1+e_3\over(1+e_2)^2}}\nonumber\\
&\approx s_3(1+e_3-2e_2+3e_2^2-2e_2e_3+\dots)\\
\left(s_3^{\rm meas.}\right)^2&=\displaystyle{s_3^2{(1+e_3)^2\over(1+e_2)^4}}\nonumber\\
&\approx s_3^2(1+2e_3-4e_2+e_3^2-8e_2e_3+10e_2^2+\dots).
\end{eqnarray}
Of course this expansion is correct only if the sample size is larger
than the smoothing scale.

\subsection*{Results for the variance and for $s_3$}

Then when we apply the previous rules on the variables
$e_p$ and get
\begin{eqnarray}
&\displaystyle{{\mg\ovl{\kappa_{\rm meas.}^2(\theta_0)}\md-\mg\kappa^2(\theta_0)\md
\over\mg\kappa^2(\theta_0)\md}=
{\epsilon^2-\mg \kappa^2(\theta_{\rm sample})\md 
\over \mg \kappa^2(\theta_0)\md}};\\
&\displaystyle{{\Delta\mg\kappa_{\rm meas.}^2(\theta_0)\md\over 
\mg\kappa_{\rm meas.}^2(\theta_0)\md}=
\sqrt{c_{22}\ \mg\kappa^2(\theta_{\rm sample})\md}};\\
&\displaystyle{{\mg s_3^{\rm meas.}\md-s_3 \over s_3}}=\nonumber\\
&\displaystyle{\left(2-3{c_{21}\over s_3}\right)\ 
{\mg\kappa^2(\theta_{\rm sample})\md \over \mg\kappa^2(\theta_0)\md}-
2\ {\epsilon^2\over\mg\kappa^2(\theta_0)\md}};\\
&\displaystyle{{\Delta s_3^{\rm meas.}\over s_3^{\rm meas.}}}=\nonumber\\
&\displaystyle{\sqrt{ \left(
{c_{33}\over s_3^2}+4 c_{22}-4{c_{32}\over s_3}\right)\ 
\mg\kappa^2(\theta_{\rm sample})\md}}.
\end{eqnarray}
One can distinguish two different effects of the cosmic variance for
the uncertainties of the results. There is first a systematic shift
proportional to the ratio of the variances. And secondly a cosmic
scatter is expected to appear, which is directly proportional
to the RMS of the fluctuations of the convergence
at the sample scale. It is thus crucial to have a catalogue
as large as possible!

Using the numerical  results of the previous appendix we have
\begin{eqnarray}
&\displaystyle{{\mg\ovl{\kappa_{\rm meas.}^2(\theta_0)}\md-\mg\kappa^2(\theta_0)\md
\over\mg\kappa^2(\theta_0)\md}=-0.13};\\
&\displaystyle{{\Delta\overline{\kappa_{\rm meas.}^2(\theta_0)}\over 
\mg\kappa_{\rm meas.}^2(\theta_0)\md}=0.14};\\
&\displaystyle{{\mg s_3^{\rm meas.}\md-s_3 \over s_3}=-0.07};\\
&\displaystyle{{\Delta s_3^{\rm meas.}\over s_3^{\rm meas.}}=0.04.}
\end{eqnarray}
The cosmic errors of course decreases with the size of the sample.
What the previous results show is that the errors are 
already quite small when the size of the sample is $25\ deg^2$ 
and would allow, for a perfectly well known source distribution,
a determination of the cosmological parameters at the 5 to 10\%
level. Note that $s_3$ appears to be less sensitive to the cosmic
variance because it is a ratio of moments.

\end{document}